# Machine learning-accelerated search of superconductors in B-C-N based compounds and $R_3Ni_2O_7$-type nickelates

Xiaoying Li,[1,2] Wenqian Tu,[1,2] Run Lv,[1,2] Li'e Liu,[1,2] Dingfu Shao,[1] Yuping Sun,[3,1,4] and Wenjian Lu[1,*]

[1]Key Laboratory of Materials Physics, Institute of Solid State Physics, HFIPS, Chinese Academy of Sciences, Hefei 230031, China

[2]University of Science and Technology of China, Hefei 230026, China

[3]High Magnetic Field Laboratory, HFIPS, Chinese Academy of Sciences, Hefei 230031, China

[4]Collaborative Innovation Center of Microstructures, Nanjing University, Nanjing 210093, China

Superconductor research has traditionally depended on experiments and theoretical approaches. However, the rapid advancement of data-driven methods and machine learning (ML) has opened avenues for accelerating superconductor discovery. Here, we integrated ML with density functional theory (DFT) calculations to efficiently screen conventional B-C-N based superconductors and identify potential high-$T_C$ candidates among $R_3Ni_2O_7$-type bilayer nickelates. We identified 12 new binary and ternary B-C-N based superconductors with $T_C \geq 10$ K, including 3 with $T_C \geq 25$ K, such as two structural forms of $B_2CN$ ($T_C$ = 44.8 K and 41.5 K) and $TiNbN_2$ ($T_C$ = 26.2 K). These materials share a common feature of strong $\sigma$-bonds, which is key to achieving relatively high $T_C$. Moreover, we proposed $Tb_3Ni_2O_7$ ($T_C$ = 61.6 K) and $Ac_3Ni_2O_7$ ($T_C$ = 70.3 K) as potential high-$T_C$ nickelate superconductors under high pressure. Their electronic structures closely resemble those of $La_3Ni_2O_7$, especially in the hole-type band dominated by Ni-3$d_{z^2}$ orbital character. We also analyzed feature importance in the ML results for both conventional and high-$T_C$ superconductors. These results advance the search for new superconductors and enhance the fundamental understanding of superconducting mechanisms.

*Corresponding author: wjlu@issp.ac.cn

## 1. Introduction

Since the discovery of superconductivity in mercury (Hg) by Onnes in 1911 [1], the pursuit for materials with high critical temperatures ($T_C$) has become a hot topic in condensed matter physics. Conventional superconductivity is explained by the Bardeen-Cooper-Schrieffer (BCS) theory [2], which proposes that light atomic masses can lead to a higher Debye frequency and thus enhance $T_C$. A prominent example is $MgB_2$, which holds the record $T_C$ of 39 K among conventional superconductors [3]. In recent decades, $T_C$ values have risen dramatically with the discovery of cuprate, iron-based, and nickelate superconductors, many exceeding the liquid nitrogen boiling temperature (77 K). Notably, the recent observation of superconductivity at 80 K in $La_3Ni_2O_7$ under pressure [4] has attracted broad interest. While the square Ni-O layers in $La_3Ni_2O_7$ resemble the Cu-O planes in cuprates, a key difference lies in the bonding of nickel atoms between adjacent layers via apical oxygen [5]. The intricate electronic correlations and layered crystal structures in these high-$T_C$ systems pose significant challenges for understanding superconducting mechanisms and discovering new superconductors.

In recent years, materials science has stepped into its fourth paradigm: "Data-Based Materials Science" [6], opening new avenues for accelerating materials discovery. Traditional approaches, which rely heavily on experimental trial-and-error and theoretical calculations [7], are often slow, computationally expensive, and inefficient for high-throughput screening. In contrast, data-driven and machine learning (ML) methods offer powerful tools for predicting and exploring superconducting materials. For instance, data-guided design led to the identification of $MgB_2$-like superconductors, later confirmed experimentally [8]. Several studies have used ML regression methods to predict $T_C$ of new compounds [9,10], and more methods, such as convolutional neural networks (CNNs), have been employed to improve accuracy, including unconventional transition metal oxides [11].

Nevertheless, many such predictions remain purely numerical estimates of $T_C$, lacking mechanistic insight or experimental validation. To address these limitations, researchers have developed integrated frameworks combining ML with density functional theory (DFT) methods to enhance the reliability of predictions for conventional superconductors. For example, Hoffmann *et al.* systematically investigated antiperovskites [12] and full-Heusler compounds [13], identifying multiple correlated superconductors. Similar approaches have been applied to high-pressure hydride systems [14—17]. These investigations demonstrate the potential of ML not only for accelerating discovery but also for mapping uncharted regions of the chemical space.

In this work, we combine ML and DFT calculations to explore the superconducting potential of binary and ternary B-C-N based compounds and the bilayer nickelates $R_3Ni_2O_7$ (R = rare-earth elements). We identified 12 new binary and ternary B-C-N based conventional superconductors with $T_C > 10$ K, including two structural forms of $B_2CN$ ($T_C = 44.8$ K and 41.5 K) and $TiNbN_2$ ($T_C = 26.2$ K). These materials exhibit strong $\sigma$-bonding characteristics, similar to those driving superconductivity in $MgB_2$. Our ML model also predicts $Tb_3Ni_2O_7$ and $Ac_3Ni_2O_7$ as potential superconducting candidates, with estimated $T_C$ values of 61.6 K and 70.3 K at 30 GPa, respectively. Electronic structure analysis reveals that the hole-type bands contributed by Ni-3$d_{z^2}$ orbital in $Tb_3Ni_2O_7$ and $Ac_3Ni_2O_7$ resemble those in $La_3Ni_2O_7$, with a similar hole concentration ($n_{\text{hole}} \sim 0.135$). By interpreting key features identified by the ML model, we gain deeper insight into the mechanisms underlying both conventional and high-$T_C$ superconductivity. This work not only demonstrates how integrated ML-DFT frameworks can simultaneously achieve reliable predictions of superconductors but also deepens our understanding of their fundamental mechanisms, thereby bridging the gap between ML screening and physical insight.

## 2. Methodology

### *2.1 Workflow*

Figure 1 presents our workflow, which integrates ML methods and DFT calculations, enabling the efficient identification of superconducting candidates. The specific details of this workflow will be explained in the subsequent paragraphs.

### *2.2 Learning data preparation*

High-quality learning data is essential for developing robust ML models. In this study, we utilized the SuperCon database [18], a rigorously curated and maintained resource by the National Institute for Materials Science (NIMS) of Japan. This database systematically archives experimentally validated superconductors and their associated literature. Our learning dataset includes 5870 conventional superconductors and 3940 cuprate high-temperature superconductors, with only two key attributes retained for each compound: the chemical formula and experimental $T_C$. The dataset exhibits a broad $T_C$ range: 0—41.4 K for conventional superconductors and 0—143.0 K for cuprates. Introducing diverse superconducting materials in training improves the accuracy and generalization of the ML models.

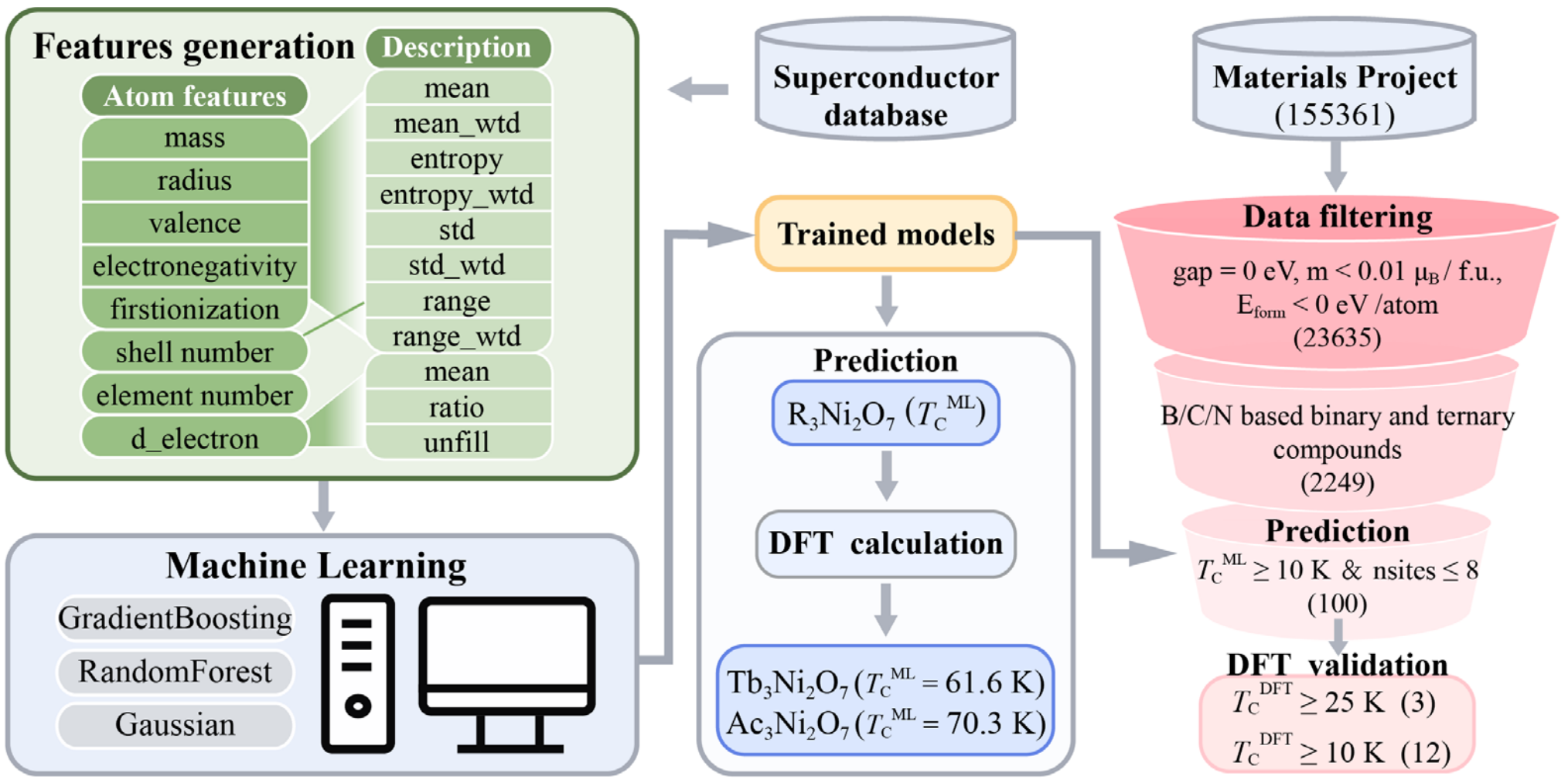

FIG. 1. Workflow for accelerating discovery of potential superconductors. The workflow involves: (1) Generating features for experimentally discovered superconductors as learning data, (2) Training ML models, (3) Filtering candidate materials based on ML-predicted $T_C$, and (4) Validating predictions by DFT calculations.

The foundation of feature generation lies in selecting physically interpretable elemental properties. We first created eight fundamental atomic attributes for each element: number of elements, atomic mass, atomic radius, valence, electronegativity, first ionization energy, number of electron shells, and number of *d* electrons. These properties have a critical influence on bonding dynamics and electronic interactions within crystals. For instance, atomic mass and radius affect lattice vibrations (phonons), while valence electrons, electronegativity, and first ionization energy modulate charge transfer. To connect atomic-scale properties with macroscopic behavior, we followed a previous study [19] and applied a series of mathematical operations to these properties, as detailed in Tab. I. Through systematic combinations, each chemical formula possesses 45 distinct features. The transformed feature space offers two key advantages for ML models: it captures interactions among elements in complex bonding networks, reflecting real material behavior, and provides richer, more comprehensive material-specific information for enhanced learning.

TABLE I. Summaries of the procedure for feature generation from the material's chemical formula. The last column serves as an example: operations are performed on the features based on the atomic mass extracted for $MgB_2$. Mg and B's atomic mass values are $t_1 = 24.305$ and $t_2 = 10.811$ respectively. Here: $p_1 = \frac{1}{1+2} = \frac{1}{3}$, $p_2 = \frac{2}{1+2} = \frac{2}{3}$, $w_1 = \frac{t_1}{t_1+t_2} \approx 0.692$, $w_2 = \frac{t_2}{t_1+t_2} \approx 0.308$, $A = \frac{p_1 w_1}{p_1 w_1 + p_2 w_2} \approx 0.529$, $B = \frac{p_2 w_2}{p_1 w_1 + p_2 w_2} \approx 0.471$.

| Description | Symbol | Formula | Sample value |
|---|---|---|---|
| Mean | mean | $\mu = (t_1 + t_2)/2$ | 17.558 |
| Weighted mean | mean_wtd | $\nu = (p_1 t_1 + p_2 t_2)$ | 15.309 |
| Entropy | entropy | $-w_1 \ln(w_1) - w_2 \ln(w_2)$ | 0.617 |
| Weighted entropy | entropy_wtd | $-A \ln(A) - B \ln(B)$ | 0.691 |
| Standard deviation | std | $[1/2((t_1 - \mu)^2 + (t_2 - \mu)^2)]^{1/2}$ | 6.747 |
| Weighted standard deviation | std_wtd | $[p_1(t_1 - \nu)^2 + p_2(t_2 - \nu)^2]^{1/2}$ | 6.361 |
| Range | range | $t_1 - t_2\ (t_1 > t_2)$ | 13.494 |
| Weighted range | range_wtd | $p_1 t_1 - p_2 t_2$ | 0.894 |

*2.3 DFT calculation details*

Electronic property calculations based on DFT were implemented via the Vienna *Ab initio* Simulation Package (VASP) [20,21]. The generalized gradient approximation (GGA) of the Perdew-Burke-Ernzerhof (PBE) form was employed as the exchange-correlation functional. A fine ***k***-point mesh with a resolution of 0.02 $Å^{-1}$ was used for the self-consistent-field (SCF) calculations, while a denser mesh with a resolution of 0.005 $Å^{-1}$ was employed for Fermi surface (FS) calculations. The projector augmented-wave (PAW) method with a 600 eV plane-wave kinetic cut-off energy was used. The phonon and electron-phonon coupling (EPC) calculations were performed by using the QUANTUM ESPRESSO (QE) package [22] with norm-conserving pseudopotentials (NCPP). The plane-wave kinetic energy cutoff and charge density were set as 60 and 600 Ry, respectively. To better describe the strong correlation of Ni-3*d* electrons in bilayer nickelates $R_3Ni_2O_7$, we use the DFT+U method, with Hubbard U = 3 eV and Hund's parameter J = 0.4 eV, also adopted in a previous study [23]. The 4*f* electrons of rare-earth R atoms were considered as core electrons. For conventional superconductors, we use the Allen-Dynes-modified McMillan equation [24]: $T_C = \frac{\omega_{log}}{1.2} exp\left[\frac{-1.04(1+\lambda)}{\lambda(1-0.62\mu^*)-\mu^*}\right]$, where EPC strength $\lambda = \sum_{qv} \lambda_{qv} \int \frac{\alpha^2 F(\omega)}{\omega} d\omega$ and Eliashberg spectral function $\alpha^2 F(\omega) = \frac{1}{2\pi N(E_F)} \sum_{qv} \delta(\omega - \omega_{qv}) \frac{\gamma_{qv}}{h\omega_{qv}}$, to quantitatively evaluate $T_C$.

## 3. Results and discussion

*3.1 Model training*

Using the prepared superconductor datasets with 45 distinct features, we trained three ML models — Gradient Boosting, Random Forest, and Gaussian Regression models — to evaluate their performance in estimating the $T_C$ of superconductors. Each model has unique characteristics and strengths. We rigorously compared their performance using three critical parameters: mean absolute error (MAE), root mean squared error (RMSE), and coefficient of determination ($R^2$). The corresponding calculation formulas are shown in *Eqs*. (1—3). This comprehensive evaluation enables objective identification of the most appropriate model.

$$\mathrm{MAE} = \frac{1}{m}\sum_{i=1}^{m} |y_i - \hat{y}_i| \quad (1)$$

$$\mathrm{RMSE} = \sqrt{\frac{1}{m}\sum_{i=1}^{m} (y_i - \hat{y}_i)^2} \quad (2)$$

$$R^2 = 1 - \frac{\sum_{i=1}^{m} (y_i - \hat{y}_i)^2}{\sum_{i=1}^{m} (y_i - \bar{y}_i)^2} \quad (3)$$

The dataset, comprising 5870 conventional superconducting materials, was divided into a training set (90%) and a test set (10%) following the standard protocol. Three ML models were trained and evaluated on the dataset. The training results show that the Gradient Boosting model outperformed the others, achieving MAE = 1.128, RMSE = 2.139, and $R^2$ = 0.919 on the testing set (Fig. 2(a)). The evaluation results, as shown in Fig. 2(b), reveal the model's predictive capability and generalization stability, with $R^2$ (train) = 0.99 and $R^2$ (test) = 0.92.

To better investigate nickelates, we constructed a hybrid dataset integrating 3940

cuprates from the SuperCon database and 126 nickelate materials (containing different reported $T_C$ values for $LaNiO_2$, $La_3Ni_2O_7$, and $La_4Ni_3O_{10}$-type materials under various pressures) from the literature. A pressure variable was added as a new feature alongside the existing 45 features. Among the three models employed in training, the Gradient Boosting model also demonstrated the best generalization performance, with $R^2$ (train) = 0.96 and $R^2$ (test) = 0.83 (Fig. 2(c)). Consequently, the model provided robust support for advancing the in-depth investigation of conventional and high-$T_C$ superconducting materials.

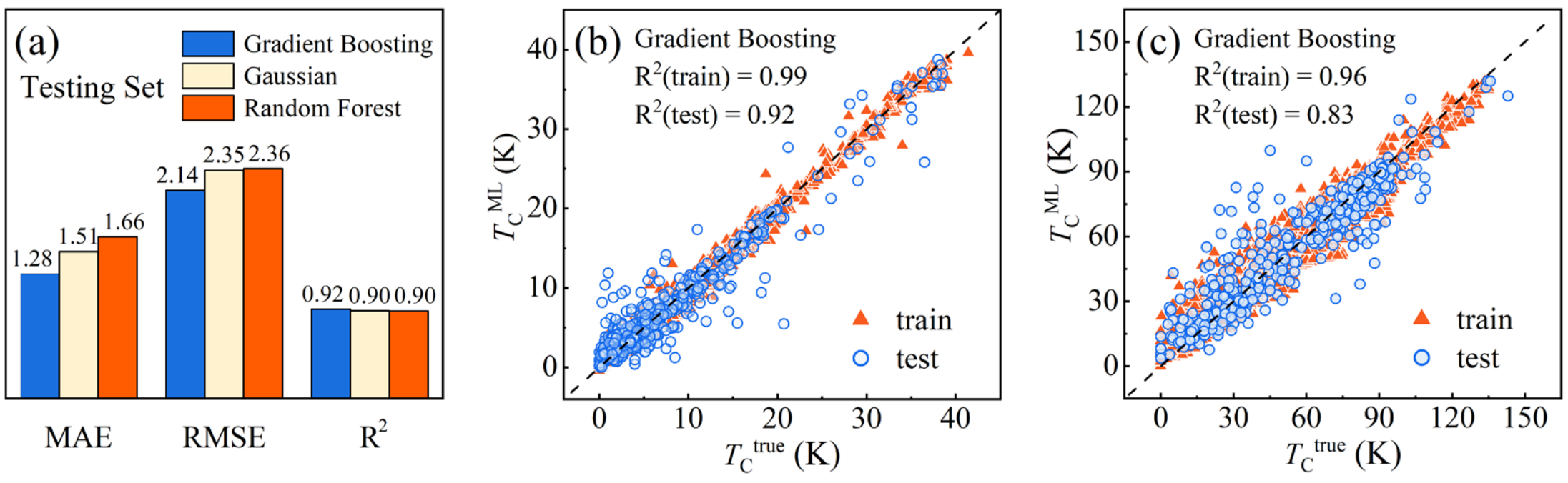

FIG. 2. Evaluation of ML models. (a) Blue, yellow, and orange denote Gradient Boosting, Gaussian, and Random Forest regression models for training conventional superconductors, respectively. (b) and (c) Benchmarking of Gradient Boosting regression models for predicting $T_C$ of conventional superconductors and high-$T_C$ superconductors, respectively. The orange triangles and blue circles represent the training set and test set data, respectively.

To further validate the prediction accuracy of our model, we applied the trained model to predict the $T_C$ of several recently experimentally reported superconductors (not included in the learning dataset). As listed in Tab. II, the ML-predicted $T_C$ ($T_C^{ML}$) values show good agreement with the experimentally observed $T_C$ values ($T_C^{exp.}$). Such consistency indicates that our trained model can deliver accurate predictions, even for materials not included in the training dataset.

TABLE II. Comparisons of ML-predicted $T_C$ and experimentally observed $T_C$ of recently discovered superconductors (not included in learning data).

| No. | Formula | $T_C^{ML}$ (K) | $T_C^{exp.}$ (K) | Ref. |
|---|---|---|---|---|
| 1 | $V_2AlN$ | 9.3 | 15.9 | [25] |
| 2 | $Ta_2V_{3.1}Si_{0.9}$ | 6.9 | 7.5 | [26] |
| 3 | $Nb_3P_2Se_2$ | 6.9 | 6.8 | [27] |
| 4 | $La_2IRu_2$ | 6.2 | 4.8 | [28] |
| 5 | $W_3Al_2C$ | 6.2 | 7.6 | [29] |
| 6 | $ZrP_{1.27}Se_{0.73}$ | 6.0 | 7.1 | [30] |
| 7 | $Nb_2PSe_2$ | 5.8 | 6.9 | [27] |
| 8 | $InNbS_2$ | 5.4 | 6.0 | [31] |
| 9 | $TlBi_2$ | 5.4 | 6.2 | [32] |
| 10 | NbReSi | 4.7 | 6.5 | [33] |
| 11 | $Mo_3ReRuC$ | 4.5 | 7.7 | [34] |
| 12 | $ThRu_3Si_2$ | 4.1 | 3.8 | [35] |

### *3.2 Feature importance discussion*

Feature importance analysis in Fig. 3(a) identifies that weighted mean mass (mass_mean_wtd), weighted mean valence (valence_mean_wtd), and weighted mean firstionization energy (firstionization_mean_wtd) are the three most critical variables for conventional superconductors, consistent with established superconductivity theory and prior experimental observations. As is well known, conventional superconductivity relies on atom mass (reflected by mass_mean_wtd) to modulate lattice vibration frequencies, thereby regulating EPC strength and ultimately determining the $T_C$. As proposed by Reynolds *et al.*, the empirical relationship demonstrated that $T_C \propto 1/\sqrt{m}$ [36]. As shown in Fig. 3(b), the weighted average attribute of mass_mean_wtd effectively captures the combined influence of different atoms in the compound on lattice vibrations.

The second key feature, valence_mean_wtd, corresponds to the valence electron concentration (VEC) [37]: $\mathrm{VEC} = (me_C + ne_A) / (m + n)$, where $m$, $n$ represent the number of cations and anions, and $e_C$, $e_A$ represent the number of the cation and anion valence electrons. VEC directly influences the density of charge carriers available for pairing and thus the magnitude of the superconducting energy gap. This relationship echoes the empirical finding first discovered by Matthias more than 60 years ago [38]. It revealed that for early discovered superconductors, two maxima in $T_C$ were found for VEC near 5 or 7. Later, $MgB_2$-related superconductors were found and associated with a $T_C$ peak for VEC near 3. Our analysis for valence_mean_wtd in Fig. 3(c) shows that there are optimal VEC values for distinct families — approximately 3 for $MgB_2$-related compounds, 4 for doped $C_{60}$, and 5 for bismuthates. While A15-family materials exhibit double peaks in their $T_C$ at VECs of ~ 4.5 and 6.5, respectively.

First ionization energy, quantifying the energy required to remove the first electron, can significantly influence a material's electronic configuration, conductivity, and ultimately its superconducting properties. Although there was no significant linear correlation between firstionization_mean_wtd and $T_C$ overall, specific values are notable in particular systems — approximately 8 for $MgB_2$-related compounds, 11 for doped $C_{60}$, 10.5 for bismuthates, and 7 for A15-family materials.

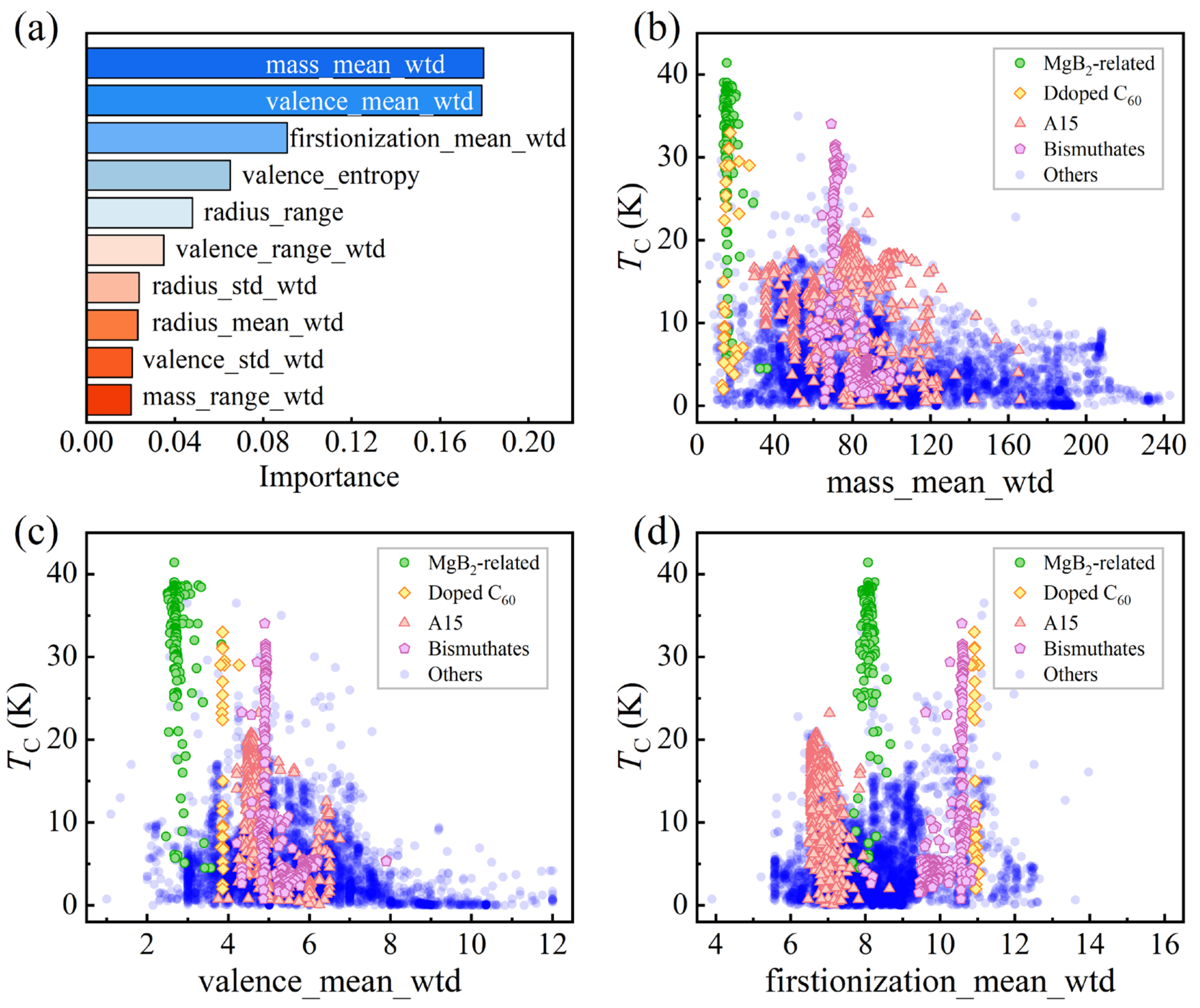

FIG. 3. (a) Top ten influential features for conventional superconductors identified by the ML model. $T_C$ is plotted vs. the three most critical (b) weighted mean mass (mass_mean_wtd), (c) weighted mean valence (valence_mean_wtd), and (d) weighted mean firstionization energy (firstionization_mean_wtd) for conventional superconductors in specific families.

We further analyze the features for high-$T_C$ superconductors and show the results in Fig. 4. The weighted standard deviation of valence electrons (valence_std_wtd) has the highest weight, indicating that the dispersion of valence electron distributions within compounds plays a dominant role in regulating $T_C$. Other important features include the weighted standard deviation of atomic radius (radius_std_wtd) and the number of unfilled *d*-orbital electrons (*d*_unfill). Figures 4(b─d) illustrate how the three key features correlate with $T_C$ across diverse material systems.

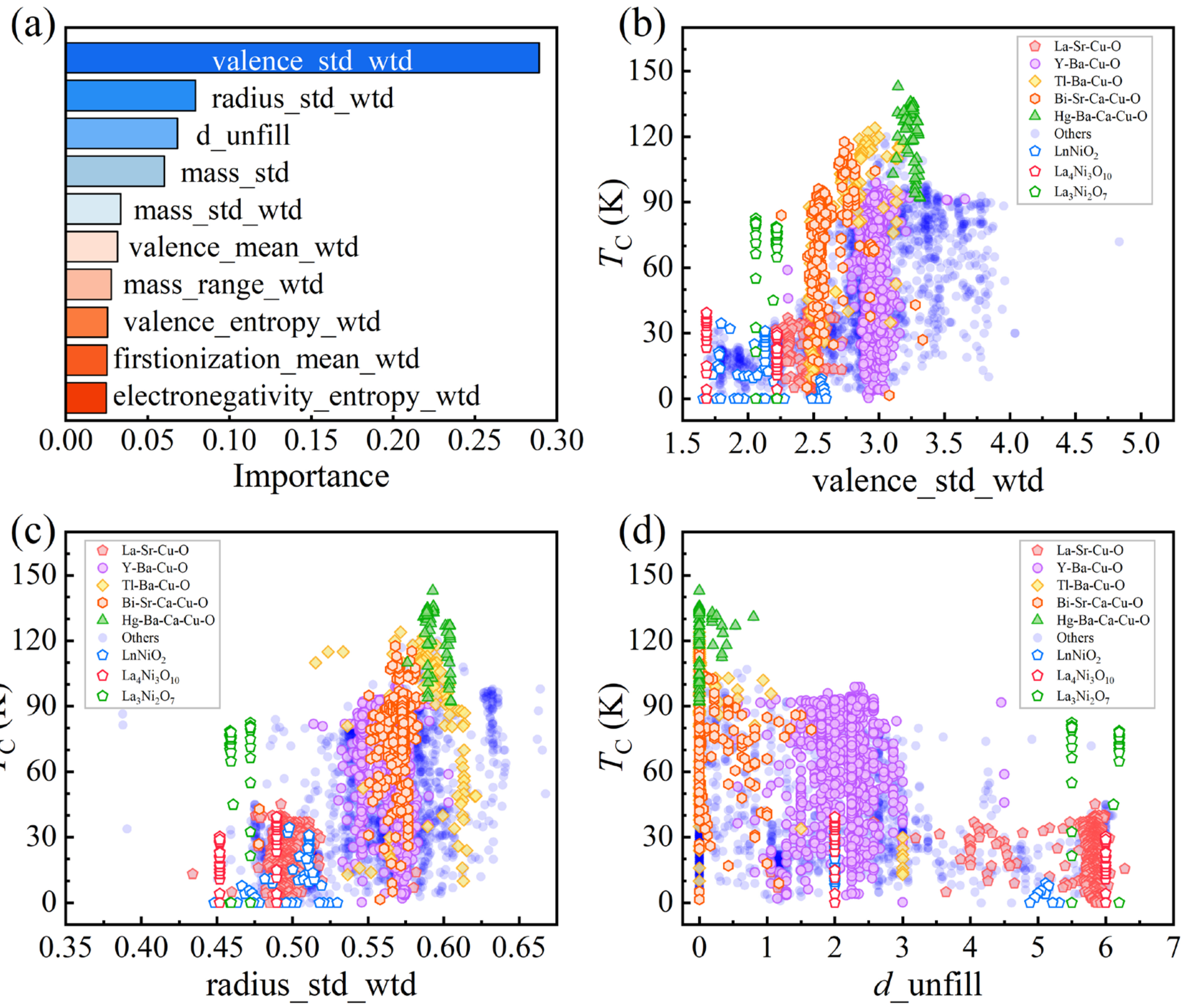

FIG. 4. (a) Top ten influential features for high-$T_C$ superconductors identified by the ML model. $T_C$ is plotted vs. the three most important (b) weighted standard deviation of valence electrons (valence_std_wtd), (c) weighted standard deviation of atomic radius (radius_std_wtd), and (d) number of unfilled *d*-orbital electrons (*d*_unfill) for cuprate and nickelate superconductors in specific families.

Figure 4(b) reveals a non-monotonic relationship of $T_C$ vs. valence_std_wtd: $T_C$ initially increases with rising valence_std_wtd, peaks at valence_std_wtd ≈ 3.25, and then declines. Despite the evident overall trend, significant scatter persists among the data points. This variability indicates that though valence_std_wtd is a primary driver of $T_C$ variations, the chemical environment — including factors such as lattice distortions, and interatomic interactions — may further refine the relationship between valence electron dispersion and $T_C$.

The weighted standard deviation of radius (radius_std_wtd) feature quantifies the statistical distribution of atomic radii within a compound, reflecting the degree of atomic-scale disorder in the crystal lattice. As shown in Fig. 4(c), the impact of radius_std_wtd on $T_C$ exhibits a strikingly similar trend to that of valence_std_wtd, peaking at approximately 0.6. This suggests that moderate variations in atomic size facilitate the emergence of high-$T_C$. For the unfilled *d*-electrons (*d*_unfill) feature in Fig. 4(d), $T_C$ basically decreases as the value increases, consistent with a prior study [39]. It exhibits the highest $T_C$ when *d*_unfill is in the range 0–1, correlating with the proportion of Cu-O layers: compounds with high Cu-O contents are near this range, while the doped systems with elements like La or Y (which have fewer *d*-electrons) increase the average *d*_unfill value due to their abundant unfilled *d*-orbitals.

### *3.3 ML predictions and DFT calculations*

As the material design aims to unearth novel superconducting materials from the vast chemical landscape, we focused on the publicly accessible Materials Project database [40]. The database contains approximately 155,361 compounds, each characterized by a wealth of computational properties, such as electronic energy bands, formation energies, and magnetic properties. Such a comprehensive dataset lays a solid foundation for the efficient screening of prospective superconductors. Our methodology employed a rigorous multi-stage screening protocol, as illustrated in Fig. 1. The screening criteria were as follows:

In the first step, we set the following screening criteria: band gap = 0 eV, magnetic moment (m) < 0.01 $\mu$B/f.u., and formation energy ($E_{\mathrm{form}}$) < 0 eV/atom. It ensures that the candidate materials have metallic, nonmagnetic properties and thermodynamic stability. Based on the above criteria, we identified 23,635 candidate materials that meet the basic requirements for conventional superconductors. In the next step, we selected the binary and ternary compounds containing light B/C/N elements, leaving the candidate materials to 2,249. Furthermore, binary and ternary compounds with simpler structures (containing fewer than 9 atoms in the unit cell) were further evaluated for their superconductivity by calculating the EPC at the DFT level.

Finally, we employed the ML-trained model to predict the $T_C$ of the previously identified 2,249 candidates. We prioritized materials with higher $T_C^{ML}$ ($T_C^{ML} \geq 10$ K) and smaller unit-cells (Nsites ≤ 8) for more accurate DFT calculations, selecting 100 materials (see Tab. SI of Supplemental Material). As listed in Tab. III, we ultimately identified 12 compounds with $T_C^{DFT} \geq 10$ K, including 3 compounds with $T_C^{DFT} \geq 25$ K, such as two structural forms of $B_2CN$ ($T_C$ = 44.8 K and 41.5 K) and $TiNbN_2$ ($T_C$ = 26.2 K). One should note that $B_2CN$ [41], $CaB_2$ [8,42,43], and $KB_6$ [9] have also been theoretically predicted as potential superconductors in previous DFT calculations.

TABLE III. Predicted superconductor candidates with $T_C^{ML} \geq 10$ K and $T_C^{DFT} \geq 5$ K. Materials Project ID number (mp-id), formula, space group, ML-predicted $T_C^{ML}$, DFT predicted $T_C^{DFT}$, and ID number in the inorganic crystal structure database (ICSD) are shown. $T_C^{DFT}$ was obtained by using the Allen-Dynes-modified McMillan equation with typical Coulomb pseudopotential $\mu^*$ = 0.10.

| No. | mp-id | Formula | Space group | $T_C^{ML}$ (K) | $T_C^{DFT}$ (K) | ICSD |
|---|---|---|---|---|---|---|
| 1 | mp-1008525 | $B_2CN$ | R3m | 16.3 | 44.8 | 183792 |
| 2 | mp-1008526 | $B_2CN$ | P3m1 | 16.3 | 41.5 | 183791 |
| 3 | mp-35869 | $TiNbN_2$ | $I4_1/amd$ | 15.7 | 26.2 | − |
| 4 | mp-37179 | $Ta_2CN$ | $I4_1/amd$ | 9.6 | 19.7 | − |
| 5 | mp-1009695 | $CaB_2$ | P6/mmm | 9.7 | 19.5 | 186764 |
| 6 | mp-1008527 | $B_2CN$ | $P\bar{4}m2$ | 16.3 | 16.8 | 183790 |
| 7 | mp-1216616 | $V_2CN$ | $R\bar{3}m$ | 14.8 | 16.2 | − |
| 8 | mp-1215180 | $ZrTiN_2$ | P4/mmm | 12.2 | 14.2 | − |
| 9 | mp-1224257 | $HfTiN_2$ | P4/mmm | 11.7 | 13.1 | − |
| 10 | mp-1215188 | $ZrTiN_2$ | $R\bar{3}m$ | 12.2 | 11.7 | − |
| 11 | mp-1076 | $KB_6$ | $Pm\bar{3}m$ | 25.1 | 11.7 | 98990 |
| 12 | mp-1224247 | $HfZrN_2$ | $R\bar{3}m$ | 12.7 | 10.8 | − |
| 13 | mp-1224286 | $HfTiN_2$ | $R\bar{3}m$ | 11.7 | 9.9 | − |
| 14 | mp-1220685 | $Nb_2Mo_2C_3$ | $R\bar{3}m$ | 9.7 | 9.8 | − |
| 15 | mp-1215387 | $Zr_4CN_3$ | $R\bar{3}m$ | 14.0 | 8.5 | − |
| 16 | mp-1217026 | $TiB_4Mo$ | P6/mmm | 10.9 | 8.4 | 43665 |
| 17 | mp-1217107 | $Ti_4CN_3$ | $R\bar{3}m$ | 12.9 | 7.3 | − |
| 18 | mp-1224384 | $Hf_4CN_3$ | $R\bar{3}m$ | 10.0 | 6.5 | − |

For conventional superconductors, EPC serves as the key mechanism for the formation of superconducting Cooper pairs. Here, we systematically calculate the phonon dispersion and EPC of the ML-predicted 100 B-C-N based superconducting candidates by DFT calculations. We show the results of four typical materials in Fig. 5 (see more results in Figs. S6 and S7 of Supplemental Material). These compounds' structures are dynamically stable due to the absence of imaginary (negative) phonon frequencies. When examining the Eliashberg function, we observe that the electron-phonon interactions are dominant for soft phonon modes. For example, the EPC of $MgB_2$ is predominantly distributed at higher frequencies, whereas the EPC of $B_2CN$ exhibits significant strength across both low and intermediate frequencies, resulting in its notably high $T_C$.

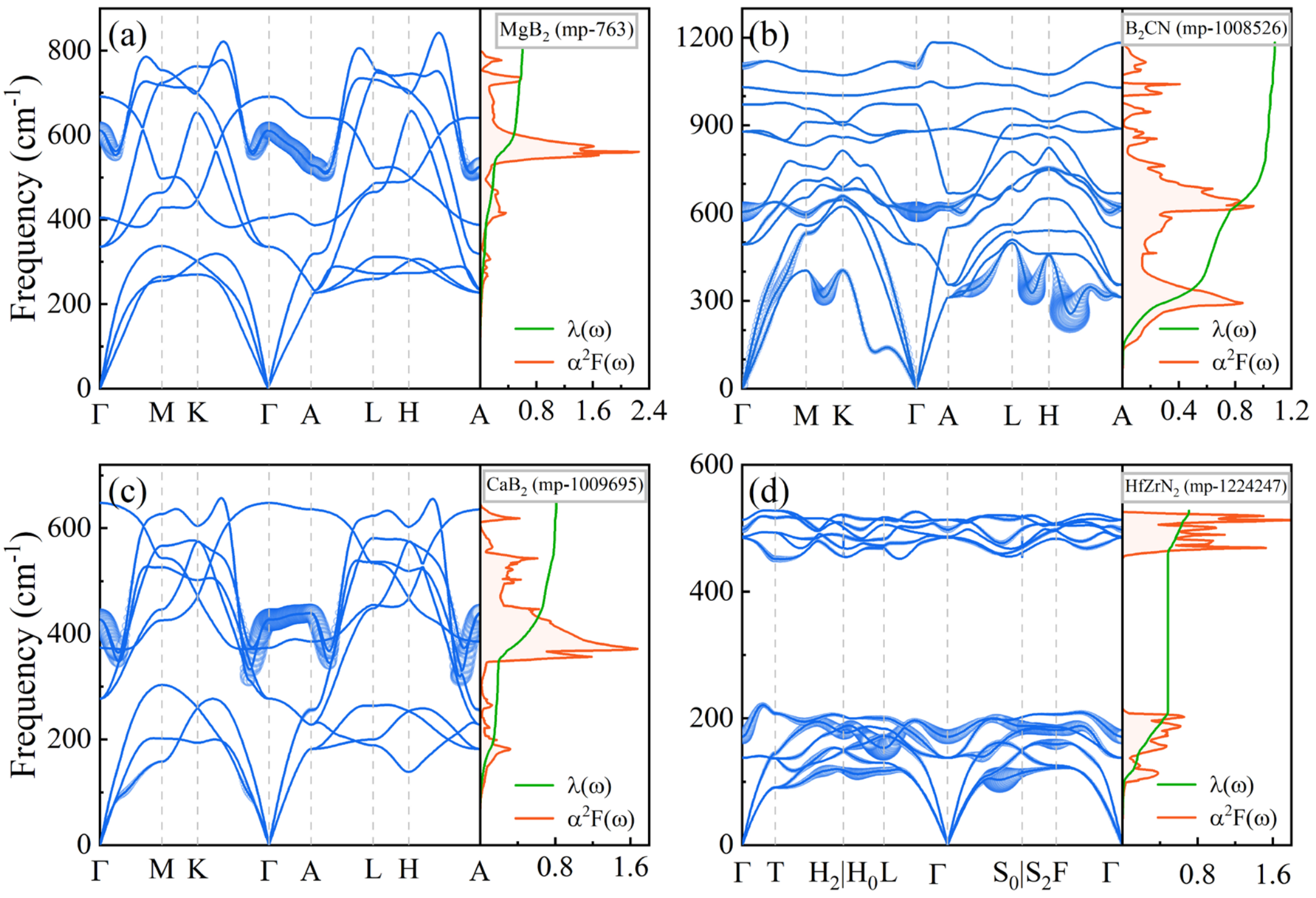

FIG. 5. Phonon spectra of (a) $MgB_2$ (mp-763), (b) $B_2CN$ (mp-1008525), (c) $CaB_2$ (mp-1009695), and (d) $HfZrN_2$ (mp-1224247) (left). The size of the blue circle is proportional to the magnitude of the electron-phonon coupling strength ($\lambda_{qv}$). The figures on the right show the Eliashberg spectral function $\alpha^2F(\omega)$ (orange line) and $\lambda(\omega)$ (green line).

We further investigated the electronic structures of the ML-predicted B-C-N based superconducting candidates. We found that most compounds share a similar feature with strong $\sigma$-bonds. For example, as illustrated in Fig. 6, the orbital-projected band structures of $MgB_2$ and $CaB_2$ show that B atoms adopt $sp^2$ hybridization, forming in-plane $\sigma$ bonds (head-on *px* and *py* orbital overlap) and vertical $\pi$ bonds (side-by-side *pz* orbital overlap). The $\sigma$-bonding bands directly cross the Fermi energy level and become the main source of carriers for superconducting pairing. While $KB_6$ and $B_2CN$ feature $\sigma$-bonding bands formed via $sp^3$ hybridization (B-B and B-C-N bonds), resulting in strong $\sigma$-bonding bands. The mechanism of metallization of strong $\sigma$-bonds can contribute to high $T_C$ values, as pointed out in previous studies [44─46].

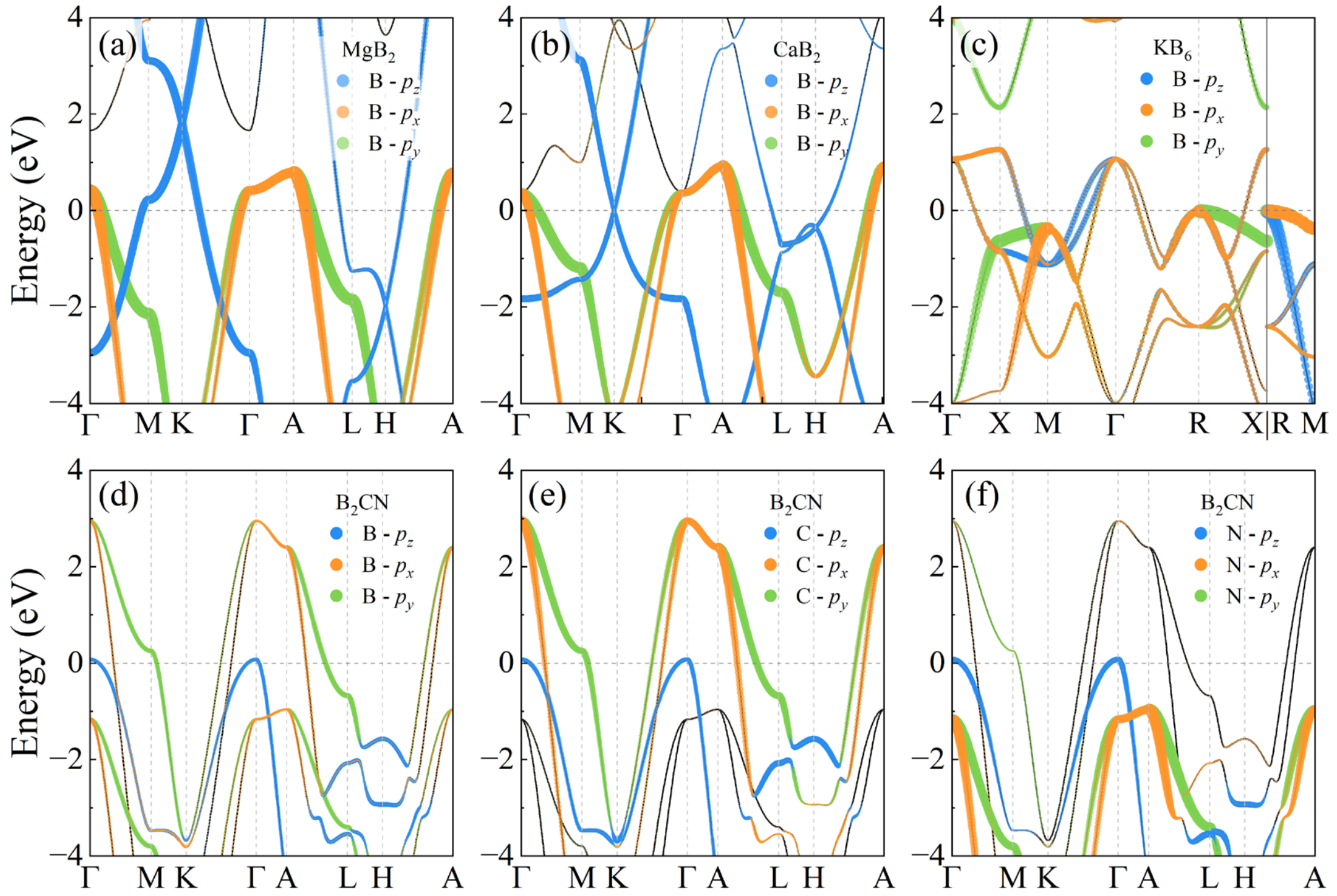

FIG. 6. Orbital-projected electronic band structures of (a) $MgB_2$, (b) $CaB_2$, (c) $KB_6$, (d─f) $B_2CN$. The widths of the blue, orange, and green lines are proportional to the weights of *pz*, *px*, and *py* orbitals, respectively.

Next, we investigated the potential high-$T_C$ candidates in $La_3Ni_2O_7$-type bilayer nickelates. Using the ML-trained model, we predicted the $T_C$ of $R_3Ni_2O_7$ (R = rare-earth elements from La to U) nickelates under a pressure of 30 GPa, at which the experimentally observed $T_C$ of $La_3Ni_2O_7$ reaches a maximum value of 80 K. As shown in Fig. 7(a), $T_C$ exhibits non-monotonic variation with increasing atomic number of R. $La_3Ni_2O_7$ demonstrates the highest $T_C$, while $Tb_3Ni_2O_7$ and $Ac_3Ni_2O_7$ show locally maximal $T_C$ values. Interestingly, our ML-identified $Tb_3Ni_2O_7$ and $Ac_3Ni_2O_7$ have also been proposed as potential high-$T_C$ superconducting nickelates in previous theoretical studies [47–49].

Given that nickelates are strongly correlated systems and belong to unconventional superconductors, whose behavior deviates from the BCS theory, one cannot directly calculate their $T_C$ by the DFT method. Instead, we adopted a comparative approach by analyzing the key features of the electronic structure of $R_3Ni_2O_7$. The DFT calculated bands and Fermi surfaces of $R_3Ni_2O_7$ are similar to those of $La_3Ni_2O_7$ (see more results in Figs. S8–S13 of Supplemental Material). As shown in Figs. 7 (b) and (d), three bands cross the Fermi energy level: two electron-type $\alpha$ and $\beta$ bands primarily contributed by the Ni-3$d_{x^2-y^2}$ orbitals and one hole-type $\gamma$ band mainly contributed by the Ni-3$d_{z^2}$ orbital. Although the band structures of $R_3Ni_2O_7$ are similar, a more detailed analysis reveals that the hole-type $\gamma$ band varies with R variation.

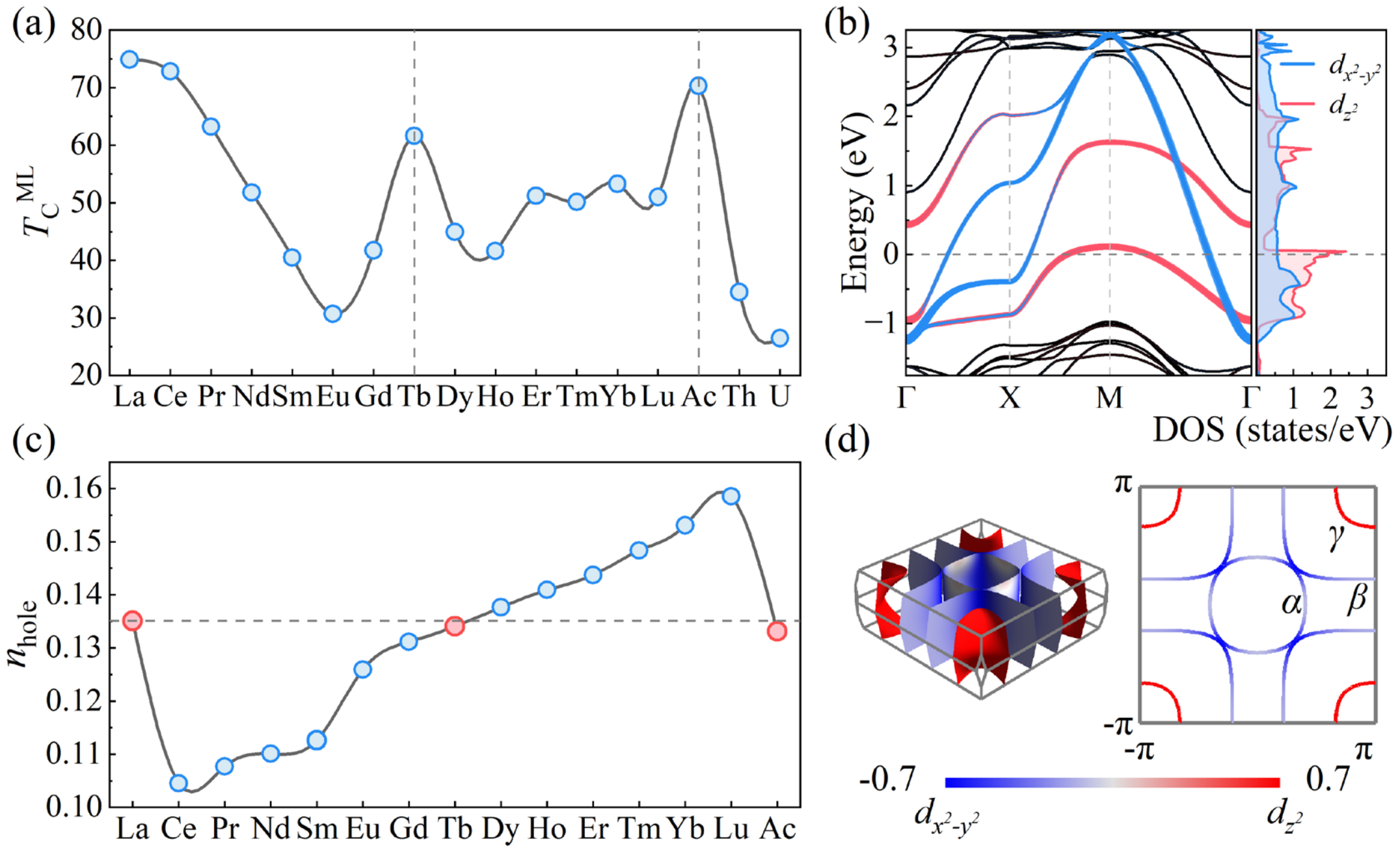

FIG. 7. (a) Predicted $T_C$ of the $R_3Ni_2O_7$ nickelates under 30 GPa pressure using the ML-trained model. $T_C$ values of $Tb_3Ni_2O_7$ and $Ac_3Ni_2O_7$ are local maxima. (b) Orbital-projected band structure and density of states of $La_3Ni_2O_7$ under 30 GPa. The widths of the red and blue lines are proportional to the weight of Ni-3$d_{z^2}$ and Ni-3$d_{x^2-y^2}$ orbitals, respectively. (c) Hole concentrations of the $\gamma$ band dominated by Ni-3$d_{z^2}$ in $R_3Ni_2O_7$. (d) Fermi surfaces in the three-dimensional Brillouin zone and projection on the two-

dimensional *kx-ky* plane with orbital weights. The three pockets are labeled as $\alpha$, $\beta$, and $\gamma$, respectively.

To quantify the variation, we calculated the hole-carrier concentrations contributed by the $\gamma$ pocket, shown in Fig. 7(c). The hole concentration exhibits a rapid decrease from La to Ce, followed by a gradual increase with rising R atomic number, and then a sharp drop upon reaching Ac. Note that for Th and U, the treatment of 4*f* electrons in pseudopotentials as valence electrons leads to poor SCF convergence in DFT calculations, yielding unreliable results and therefore not presented. Surprisingly, one can find that the hole concentrations $n_{\mathrm{hole}}$ of $\gamma$ pockets for $Tb_3Ni_2O_7$ and $Ac_3Ni_2O_7$ are 0.134 and 0.133, which are very close to the value of 0.135 for $La_3Ni_2O_7$. Previous theoretical calculations propose that the $\gamma$ pocket contributed by the Ni-3$d_{z^2}$ orbital plays a dominant role in the superconductivity of $La_3Ni_2O_7$, determining the pairing symmetry as $s^{\pm}$ [4,51–54]. The $\gamma$ bands of $Tb_3Ni_2O_7$ and $Ac_3Ni_2O_7$ identified exhibit strong consistency with those of $La_3Ni_2O_7$, suggesting their potential as bilayer nickelate superconductors.

## 4. Conclusions

In this study, we developed a large-scale strategy to accelerate the discovery of superconductors by combining data-driven ML techniques and first-principles calculations. We systematically explored the binary and ternary B-C-N based conventional superconductors as well as potential high-$T_C$ $R_3Ni_2O_7$ nickelates. From an initial pool of over 150,000 materials, our ML model identified 100 binary/ternary B-C-N based compounds with $T_C^{\mathrm{ML}} \geq 10$ K. Among these, 12 compounds were further validated using DFT, yielding $T_C^{\mathrm{DFT}} \geq 10$ K, including 3 compounds with $T_C^{\mathrm{DFT}} \geq 25$ K. The predicted B-C-N based compounds exhibit strong $\sigma$-bonding interactions, similar to those of $MgB_2$. Furthermore, our approach uncovered two promising nickelate superconducting candidates, $Tb_3Ni_2O_7$ and $Ac_3Ni_2O_7$, which demonstrate high $T_C$ values comparable to $La_3Ni_2O_7$. Computational analyses reveal that these nickelates share a hole-type band derived from Ni-3$d_{z^2}$ orbitals, accompanied by comparable hole carrier concentrations. Critical feature analysis underscores key descriptors responsible for both conventional and high-$T_C$ superconductors, facilitating not only the efficient discovery of novel materials but also deeper mechanistic insights.

This work demonstrates a robust pathway for predicting both conventional and high-$T_C$ superconducting materials and contributes to a more comprehensive understanding of their underlying mechanisms.

## References

[1] H. K. Onnes, The Resistance of Pure Mercury at Helium Temperatures, Commun. Phys. Lab. Univ. Leiden **12**, 1 (1911).

[2] J. Bardeen, L. N. Cooper, and J. R. Schrieffer, Theory of superconductivity, Phys. Rev. **108**, 1175 (1957).

[3] J. Nagamatsu, N. Nakagawa, T. Muranaka, Y. Zenitani, and J. Akimitsu, Superconductivity at 39 K in magnesium diboride, Nature **410**, 63 (2001).

[4] H. L. Sun, M. W. Huo, X. W. Hu, J. Y. Li, Z. J. Liu, Y. F. Han, L. Y. Tang, Z. Q. Mao, P. T. Yang, B. S. Wang, J. G. Cheng, D. X. Yao, G. M. Zhang, and M. Wang, Signatures of superconductivity near 80 K in a nickelate under high pressure, Nature **621**, 493 (2023).

[5] C. Park and R. L. Snyder, Structures of High-Temperature Cuprate Superconductors, J. Am. Ceram. Soc. **78**, 3171 (1995).

[6] R. Kitchin, Big data, new epistemologies and paradigm shifts, Big Data Soc. **1**, 1 (2014).

[7] G. R. Schleder, A. C. M. Padilha, C. M. Acosta, M. Costa, and A. Fazzio, From DFT to machine learning: recent approaches to materials science — a review, J. Phys. Mater. **2**, 032001 (2019).

[8] Z. Yu, T. Bo, B. Liu, Z. D. Fu, H. Wang, S. Xu, T. L. Xia, S. L. Li, and S. Meng, Superconductive materials with $MgB_2$-like structures from data-driven screening, Phys. Rev. B **105**, 214517 (2022).

[9] K. Hamidieh, A data-driven statistical model for predicting the critical temperature of a superconductor, Comput. Mater. Sci. **154**, 346 (2018).

[10] H. Gashmard, H. Shakeripour, and M. Alaei, Predicting superconducting transition temperature through advanced machine learning and innovative feature engineering, Sci. Rep. **14**, 3965 (2024).

[11] M. R. Quinn and T. M. McQueen, Identifying new classes of high temperature superconductors with convolutional neural networks, Front. Electron. Mater. **2**, 2022 (2022).

[12] N. Hoffmann, T. F. T. Cerqueira, J. Schmidt, and M. A. L. Marques, Superconductivity in antiperovskites, npj Comput. Mater. **8**, 150 (2022).

[13] N. Hoffmann, T. F. T. Cerqueira, P. Borlido, A. Sanna, J. Schmidt, and M. A. L. Marques, Searching for ductile superconducting Heusler $X_2YZ$ compounds, npj Comput. Mater. **9**, 138 (2023).

[14] L. Gu, Y. Liu, P. Chen, H. Y. Huang, N. Chen, Y. Li, T. Lookman, Y. T. Lu, and Y. T. Su, Bond sensitive graph neural networks for predicting high temperature superconductors, MGE Adv. **2**, e48 (2024).

[15] B. W. Jiang, X. S. Luo, Y. Sun, X. Zhong, J. Lv, Y. Xie, Y. M. Ma, and H. Y. Liu, Data-driven search for high-temperature superconductors in ternary hydrides under pressure, Phys. Rev. B **111**, 054505 (2025).

[16] L. L. Zha, J. J. Jiang, Y. M. Xue, Z. B. Cheng, S. W. Yao, W. J. Hu, L. Peng, T. T. Shi, J. Chen, X. L. Liu, and J. Lin, Stability and superconductivity of hexagonal prism-structured polyhydrides $X_2MgH_{18}$ (X = Li, Na, K, Rb, Cs) under moderate pressure, Phys. Chem. Chem. Phys. **27**, 3844 (2025).

[17] Y. Du, Z. F. Wang, H. Y. Liu, G. J. Liu, and X. Zhong, Superconductivity of electron-doped chalcohydrides under high pressure, Phys. Rev. Res. **7**, 013049 (2025).

[18] National Institute of Materials Science, Materials Information Station, SuperCon (2011).

[19] K. Hamidieh, A data-driven statistical model for predicting the critical temperature of a superconductor, Comput. Mater. Sci. **154**, 346 (2018).

[20] G. Kresse and J. Furthmüller, Efficient iterative schemes for *ab initio* total-energy calculations using a plane-wave basis set, Phys. Rev. B **54**, 11169 (1996).

[21] P. E. Blöchl, Projector augmented-wave method, Phys. Rev. B **50**, 17953 (1994).

[22] P. Giannozzi, S. Baroni, N. Bonini, M. Calandra, R. Car, C. Cavazzoni, D. Ceresoli, G. L. Chiarotti, M. Cococcioni, I. Dabo, A. Dal Corso, S. de Gironcoli, S. Fabris, G. Fratesi, R. Gebauer, U. Gerstmann, C. Gougoussis, A. Kokalj, M. Lazzeri, L. Martin-Samos, N. Marzari, F. Mauri, R. Mazzarello, S. Paolini, A. Pasquarello, L. Paulatto, C. Sbraccia, S. Scandolo, G. Sclauzero, A. P. Seitsonen, A. Smogunov, P. Umari, and R. M. Wentzcovitch, QUANTUM ESPRESSO: a modular and open-source software project for quantum simulations of materials, J. Phys.: Condens. Matter **21**, 395502 (2009).

[23] Z. H. Luo, X. W. Hu, M. Wang, W. Wu, and D. X. Yao, Bilayer Two-Orbital Model

of $La_3Ni_2O_7$ under Pressure, Phys. Rev. Lett. **131**, 126001 (2023).

[24] P. B. Allen and R. C. Dynes, Transition temperature of strong-coupled superconductors reanalyzed, Phys. Rev. B **12**, 905 (1975).

[25] J. Li, X. P. Cui, Y. Jin, X. Q. Yin, S. X. Cao, Z. J. Feng, and J. C. Zhang, Appearance of superconductivity at 15.9 K in layered $V_2AlN$, J. Supercond. Nov. Magn. **30**, 1 (2017).

[26] J. N. Graham, H. Liu, V. Sazgari, C. Mielke, M. Medarde, H. Luetkens, R. Khasanov, Y. Shi, and Z. Guguchia, Microscopic probing of the superconducting and normal state properties of $Ta_2V_{3.1}Si_{0.9}$ by muon spin rotation, Commun. Mater. **5**, 225 (2024).

[27] Y. J. Lee, O. Inturu, J. H. Kim, and J. S. Rhyee, Robust bulk superconductivity by giant proximity effect in Weyl semimetal-superconducting NbP/$NbSe_2$ composites, J. Appl. Phys. **132**, 123903 (2022).

[28] H. Ishikawa, U. Wedig, J. Nuss, R. K. Kremer, R. Dinnebier, M. Blankenhorn, M. Pakdaman, Y. Matsumoto, T. Takayama, K. Kitagawa, and H. Takagi, Superconductivity at 4.8 K and violation of Pauli limit in $La_2IRu_2$ comprising Ru honeycomb layer, Inorg. Chem. **58**, 19 (2019).

[29] D. Tay, T. Shang, Y. P. Qi, T. P. Ying, H. Hosono, H. R. Ott, and T. Shiroka, *s*-wave superconductivity in the noncentrosymmetric $W_3Al_2C$ superconductor: an NMR study, J. Phys. Condens. Matter **34**, 194005 (2022).

[30] K.W. Chen, G. Chappell, S. Zhang, W. Lan, T. Besara, K. Huang, D. Graf, L. Balicas, A. P. Reyes, Superconductivity in single crystals of $ZrP_{1.27}Se_{0.73}$, Phys. Rev. B **102**, 144522 (2020).

[31] B. Zheng, X. K. Feng, B. Liu, Z. F. Liu, S. S. Wang, Y. Zhang, X. Ma, Y. Luo, C. L. Wang, R. M. Li, Z. Y. Zhang, S. T. Cui, Y. L. Lu, Z. Sun, J. F. He, S. Y. A. Yang, and B. Xiang, The coexistence of superconductivity and topological order in Van der Waals $InNbS_2$, Small **20**, 5 (2024).

[32] Z. H. Yang, Z. Yang, Q. P. Su, E. D. Fang, J. H. Yang, B. Chen, H. D. Wang, J. H. Du, C. X. Wu, and M. H. Fang, Superconductivity in $TlBi_2$ with a large Kadowaki-Woods ratio, Phys. Rev. B **106**, 224501 (2022).

[33] T. Shang, D. Tay, H. Su, and T. Shiroka, Evidence of fully gapped superconductivity in NbReSi: A combined $\mu$ SR and NMR study, Phys. Rev. B **105**, 144506 (2022).

[34] Q. Q. Zhu, G. R. Xiao, W. Z. Yang, S. J. Song, G. H. Cao, and Z. Ren, $Mo_3ReRuC$:

A Noncentrosymmetric superconductor formed in the MoReRu–$Mo_2C$ System, Inorg. Chem. **61**, 43 (2022).

[35] Y. Liu, J. Li, W. Z. Yang, J. Y. Lu, B. Y. Cao, H. X. Li, W. L. Chai, S. Q. Wu, B. Z. Li, Y. L. Sun, W. H. Jiao, C. Wang, X. F. Xu, Z. Ren, and G. H. Cao, Superconductivity in Kagome metal $ThRu_3Si_2$, Chin. Phys. B **33**, 057401 (2024).

[36] M. Olsen, Superconductivity of Lead Isotopes, Nature **168**, 245 (1951).

[37] M. Pop, G. Borodi, and S. Simon, Correlation between valence electron concentration and high-temperature superconductivity, J. Phys. Chem. Solids **61**, 12 (2000).

[38] B. T. Matthias, Empirical relation between superconductivity and the number of valence electrons per atom, Phys. Rev. **97**, 74 (1955).

[39] V. Stanev, C. Oses, A. G. Kusne, E. Rodriguez, J. Paglione, S. Curtarolo, and I. Takeuchi, Machine learning modeling of superconducting critical temperature, npj Comput. Mater. **4**, 29 (2018).

[40] S. P. Ong, S. Cholia, A. Jain, M. Brafman, D. Gunter, G. Ceder, and K. A. Persson, The Materials Application Programming Interface (API): A simple, flexible, and efficient API for materials data based on REpresentational State Transfer (REST) principles, Comput. Mater. Sci. **97**, 209 (2015).

[41] X. X. Zhang, G. L. Yu, H. Chen, Y. Zhao, T. M. Cheng, Q. Li, Electron deficiency but semiconductive diamond-like $B_2CN$ originated from three-center bonds, Phys. Chem. Chem. Phys. **23**, 4 (2021).

[42] N. I. A. Ahmed and C. Parlak, First-principles calculations on mechanical and electronical properties of $AlB_2$-type $CaB_2$, Mater. Today Commun. **38**, 108290 (2024).

[43] H. J. Choi, S. G. Louie, and M. L. Cohen, Prediction of superconducting properties of $CaB_2$ using anisotropic Eliashberg theory, Phys. Rev. B **80**, 064503 (2009).

[44] M. Gao, Z. Y. Lu, and T. Xiang, Finding high-temperature superconductors by metallizing the $\sigma$-bonding electrons, Physics, **44**, 421 (2015).

[45] J. M. An and W. E. Pickett, Superconductivity of $MgB_2$: Covalent Bonds Driven Metallic, Phys. Rev. Lett. **86**, 4366 (2011).

[46] Y. Kong, O. V. Dolgov, O. Jepsen, and O. K. Andersen, Electron-phonon interaction in the normal and superconducting states of $MgB_2$, Phys. Rev. B **64**, 020501(R) (2011).

[47] B. Geisler, J. J. Hamlin, G. R. Stewart, R. G. Hennig, and P. J. Hirschfeld,

Structural transitions, octahedral rotations, and electronic properties of $A_3Ni_2O_7$ rare-earth nickelates under high pressure, npj Quantum Mater. **9**, 38 (2024).

[48] S. Q. Wu, Z. H. Yang, X. Ma, J. H. Dai, M. Shi, H. Q. Yuan, H. Q. Lin, and C. Cao, $Ac_3Ni_2O_7$ and $La_2AeNi_2O_6F$ (Ae=Sr, Ba): Benchmark Materials for Bilayer Nickelate Superconductivity, arXiv:2403.11713.

[49] L. C. Rhodes and P. Wahl, Structural routes to stabilize superconducting $La_3Ni_2O_7$ at ambient pressure, Phys. Rev. Mater. **8**, 044801 (2024).

[50] P. Li, G. D. Zhou, W. Lv, Y. Y. Li, C.M. Yue, H. L. Huang, L. Z. Xu, J. C. Shen, Y. Miao, W. H. Song, Z. H. Nie, Y. Q. Chen, H. Wang, W. Q. Chen, Y. B. Huang, Z. H. Chen, T. Qian, J. H. Lin, J. F. He, Y. J. Sun, Z. Y. Chen, and Q. K. Xue, Angle-resolved photoemission spectroscopy of superconducting $(La,Pr)_3Ni_2O_7/SrLaAlO_4$ heterostructures, Nat. Sci. Rev. nwaf205 (2025).

[51] Z. Luo, X. Hu, M. Wang, W. Wu, and D. X. Yao, Bilayer two-orbital model of $La_3Ni_2O_7$ under pressure. Phys. Rev. Lett. **131**, 126001 (2023).

[52] Q. G. Yang, D. Wang, and Q. H. Wang, Possible $s^{\pm}$ wave superconductivity in $La_3Ni_2O_7$, Phys. Rev. B **108**, L140505 (2023).

[53] Y. H. Gu, C. C. Le, Z. S. Yang, X. X. Wu, and J. P. Hu, Effective model and pairing tendency in the bilayer Ni-based superconductor $La_3Ni_2O_7$, Phys. Rev. B **111**, 174506 (2025).

[54] Z. Y. Shao, Y. B. Liu, M. Liu, and F. Yang, Band structure and pairing nature of $La_3Ni_2O_7$ thin film at ambient pressure, Phys. Rev. B **112**, 024506 (2025).

SUPPLEMENTAL MATERIAL

# Machine learning-accelerated search of superconductors in B-C-N based compounds and $R_3Ni_2O_7$-type nickelates

Xiaoying Li,[1,2] Wenqian Tu,[1,2] Run Lv,[1,2] Li'e Liu,[1,2] Dingfu Shao,[1] Yuping Sun,[3,1,4] and Wenjian Lu[1,*]

[1]Key Laboratory of Materials Physics, Institute of Solid State Physics, HFIPS, Chinese Academy of Sciences, Hefei 230031, China

[2]University of Science and Technology of China, Hefei 230026, China

[3]High Magnetic Field Laboratory, HFIPS, Chinese Academy of Sciences, Hefei 230031, China

[4]Collaborative Innovation Center of Microstructures, Nanjing University, Nanjing 210093, China

## 1. Machine learning for $T_C$ prediction: supplementary figures of model metrics and comparative benchmarks

We conduct a series of analyses, including model metrics, as shown in Fig. S1 and Fig. S2, clearly showing the performance differences of different models. Benchmarking charts of three machine learning methods for predicting superconducting $T_C$, as shown in Fig. S3 and Fig. S4, provide an important basis for in-depth analysis and comparison of the effectiveness of different models.

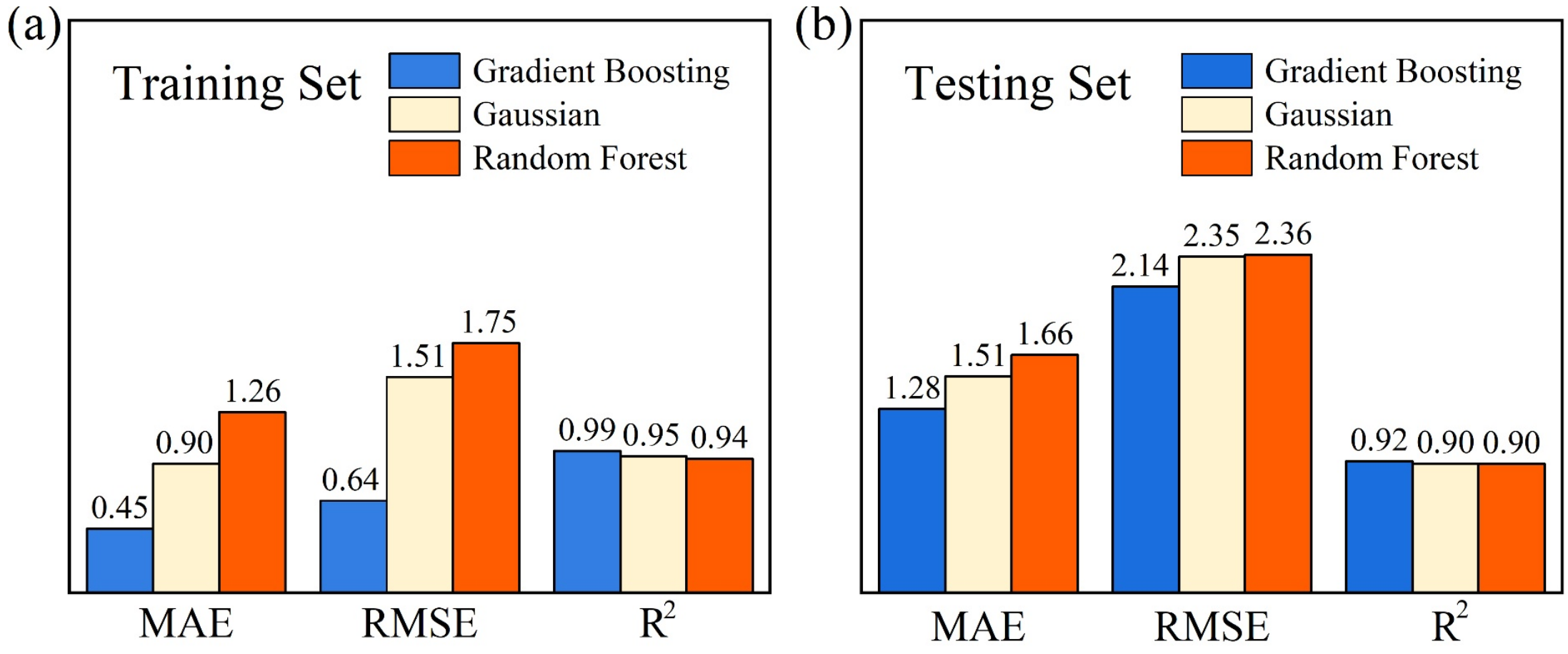

FIG. S1. The Gradient Boosting, Gaussian, and Random Forest regression models' performance in training and testing sets of conventional superconductors.

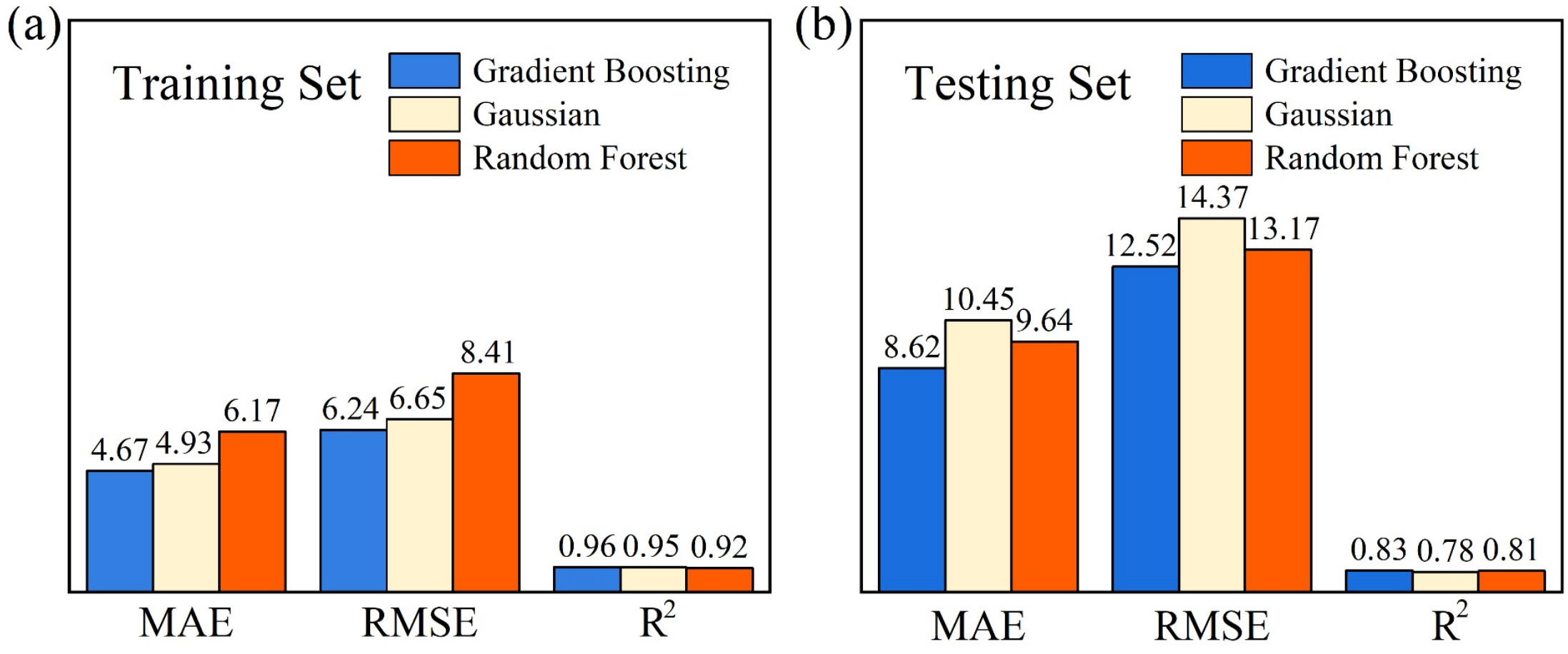

FIG. S2. The Gradient Boosting, Gaussian, and Random Forest regression models' performance in training and testing sets of cuprate and nickelate superconductors.

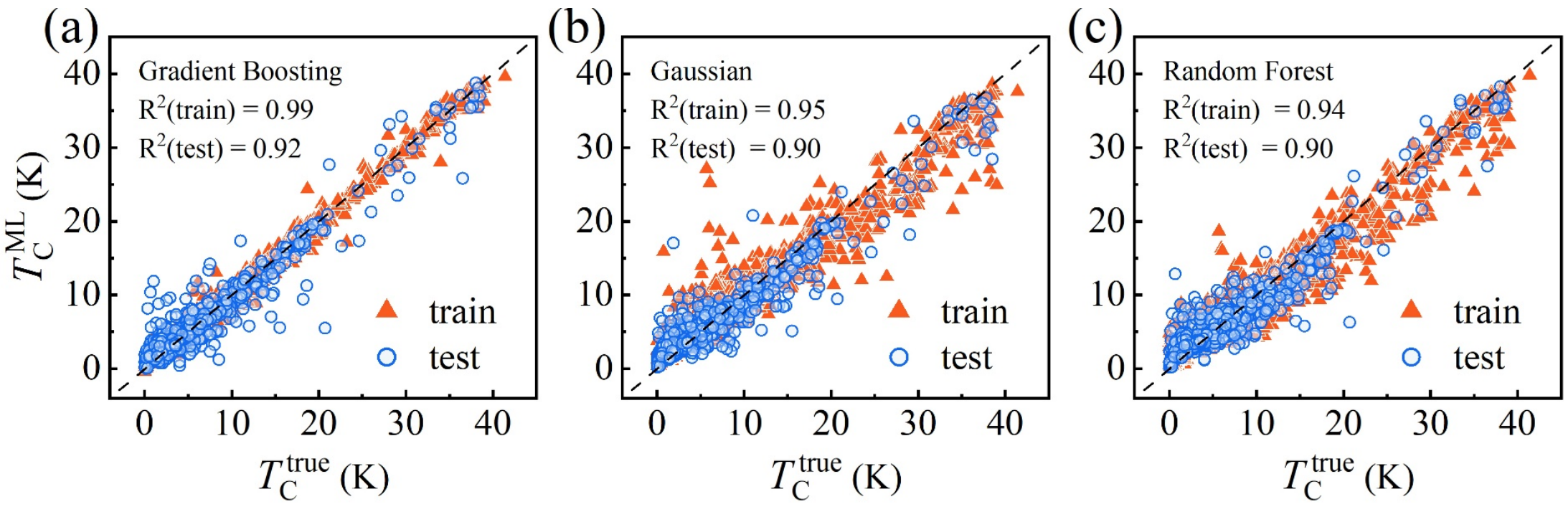

FIG. S3. Benchmarking of Gradient Boosting, Gaussian, and Random Forest regression models for predicting $T_C$ of conventional superconductors. Comparisons between true $T_C$ ($T_C^{true}$) and predicted $T_C$ ($T_C^{ML}$) for the three regression models.

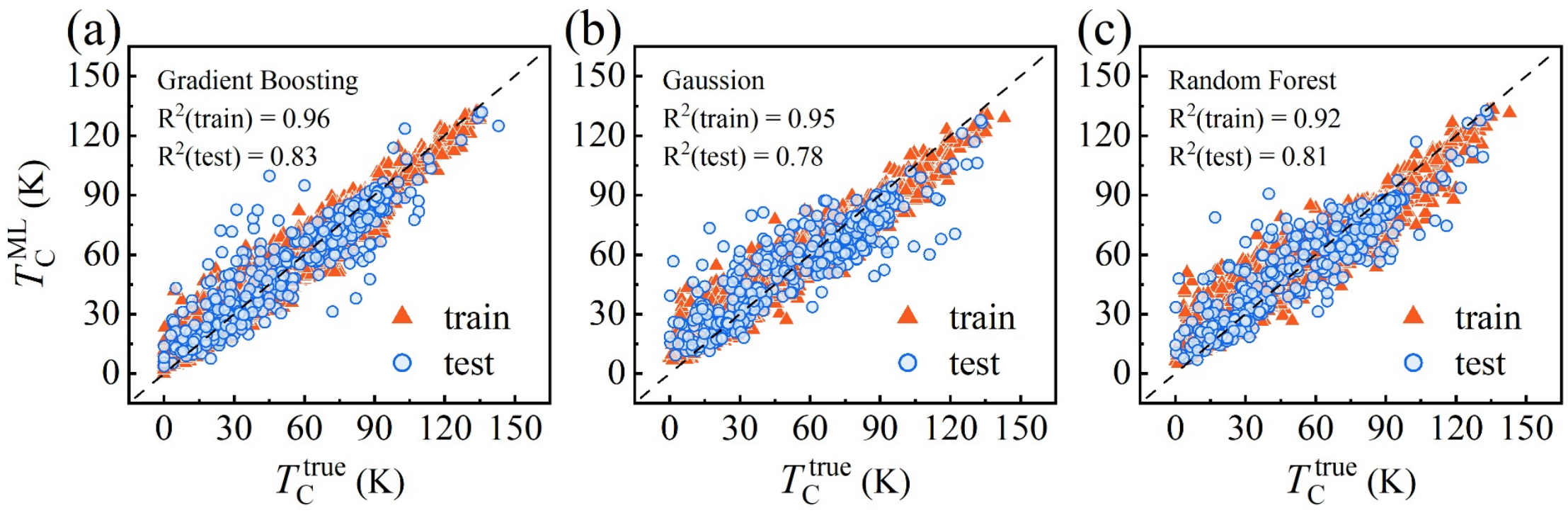

FIG. S4. Benchmarking of Gradient Boosting, Gaussian, and Random Forest regression models for predicting $T_C$ of cuprate and nickelate superconductors. Comparisons between true $T_C$ ($T_C^{true}$) and predicted $T_C$ ($T_C^{ML}$) for the three regression models.

## 2. Superconducting candidates predicted via the Gradient Boosting regression model

We illustrate the relevant information of superconducting candidates predicted via the Gradient Boosting regression model. There are total 100 compounds with $T_C^{ML} \geq$ 10 K. We have also conducted DFT calculations on these compounds. The results are shown in Table SI.

TABLE SI. Calculated superconducting candidates' properties. We present the Materials Project ID number (mp-ids), the chemical formula (Formula), the space group (Spg), the number of atoms in the primitive unit cell (NSites), the total magnetization (m in μB/f.u.), the predicted formation energy ($E_{form}$ in eV/atom), the $T_C$ predicted by machine learning ($T_C^{ML}$ in K) and the $T_C$ calculated by DFT ($T_C^{DFT}$ in K). No value of $T_C^{DFT}$ means existence of imaginary frequencies in phonon dispersion spectra, indicating structural instability.

| No. | mp-id | Formula | Spg | NSites | m | $E_{form}$ | $T_C^{ML}$ | $T_C^{DFT}$ |
|---|---|---|---|---|---|---|---|---|
| 1 | mp-1008525 | $B_2CN$ | R3m | 4 | 3.6E-03 | -0.42 | 16.3 | 44.8 |
| 2 | mp-1008526 | $B_2CN$ | P3m1 | 4 | 2.4E-03 | -0.41 | 16.3 | 41.5 |
| 3 | mp-35869 | $TiNbN_2$ | $I4_1/amd$ | 8 | 2.7E-05 | -1.50 | 15.7 | 26.2 |
| 4 | mp-37179 | $Ta_2CN$ | $I4_1/amd$ | 8 | 1.3E-04 | -0.83 | 9.6 | 19.7 |
| 5 | mp-1009695 | $CaB_2$ | P6/mmm | 3 | 4.1E-05 | -0.19 | 9.7 | 19.5 |
| 6 | mp-1008527 | $B_2CN$ | $P\bar{4}m2$ | 4 | 2.4E-03 | -0.44 | 16.3 | 16.8 |
| 7 | mp-1216616 | $V_2CN$ | $R\bar{3}m$ | 4 | 2.6E-03 | -0.80 | 14.8 | 16.2 |
| 8 | mp-999355 | NbN | $P6_3mc$ | 4 | 5.0E-04 | -0.86 | 15.0 | 15.2 |
| 9 | mp-1220725 | $Nb_2CN$ | $R\bar{3}m$ | 4 | 1.7E-03 | -0.80 | 16.3 | 14.5 |
| 10 | mp-1215180 | $ZrTiN_2$ | P4/mmm | 4 | 6.1E-04 | -1.87 | 12.2 | 14.2 |
| 11 | mp-1217924 | $TaNbC_2$ | $R\bar{3}m$ | 4 | 1.5E-06 | -0.52 | 9.7 | 13.2 |
| 12 | mp-1224257 | $HfTiN_2$ | P4/mmm | 4 | 5.3E-05 | -1.94 | 11.7 | 13.1 |
| 13 | mp-763 | $MgB_2$ | P6/mmm | 3 | 1.0E-07 | -0.18 | 35.0 | 11.8 |
| 14 | mp-1215188 | $ZrTiN_2$ | $R\bar{3}m$ | 4 | 1.6E-04 | -1.91 | 12.2 | 11.7 |
| 15 | mp-1076 | $KB_6$ | $Pm\bar{3}m$ | 7 | 0.0E+00 | -0.08 | 25.1 | 11.7 |
| 16 | mp-1224247 | $HfZrN_2$ | $R\bar{3}m$ | 4 | 1.0E-07 | -1.99 | 12.7 | 10.8 |
| 17 | mp-1224286 | $HfTiN_2$ | $R\bar{3}m$ | 4 | 6.1E-06 | -1.95 | 11.7 | 9.9 |
| 18 | mp-1220685 | $Nb_2Mo_2C_3$ | $R\bar{3}m$ | 7 | 3.2E-04 | -0.10 | 9.7 | 9.8 |
| 19 | mp-1215387 | $Zr_4CN_3$ | $R\bar{3}m$ | 8 | 1.0E-07 | -1.63 | 14.0 | 8.5 |
| 20 | mp-1217026 | $TiB_4Mo$ | P6/mmm | 6 | 5.9E-06 | -0.71 | 10.9 | 8.4 |
| 21 | mp-1217107 | $Ti_4CN_3$ | $R\bar{3}m$ | 8 | 0.0E+00 | -1.65 | 12.9 | 7.3 |
| 22 | mp-1224384 | $Hf_4CN_3$ | $R\bar{3}m$ | 8 | 0.0E+00 | -1.72 | 10.0 | 6.5 |
| 23 | mp-1079333 | $B_2CN$ | Pmma | 8 | 3.0E-05 | -0.52 | 16.3 | 4.9 |
| 24 | mp-644278 | $Y_2H_2C$ | $P\bar{3}m1$ | 5 | 0.0E+00 | -0.58 | 13.7 | 4.9 |
| 25 | mp-1009471 | NbN | $Pm\bar{3}m$ | 2 | 2.3E-05 | -0.61 | 15.0 | 4.2 |
| 26 | mp-9964 | $Ti_3TlN$ | $Pm\bar{3}m$ | 5 | 3.6E-04 | -0.85 | 11.0 | 3.9 |
| 27 | mp-999357 | NbN | $P6_3/mmc$ | 4 | 7.2E-06 | -1.09 | 15.0 | 3.9 |
| 28 | mp-1246929 | $Zr_4BN_3$ | $Pm\bar{3}m$ | 8 | 6.0E-07 | -1.69 | 12.5 | 3.4 |
| 29 | mp-2646985 | $V_2GaN$ | $P6_3/mmc$ | 8 | 3.0E-04 | -0.86 | 9.5 | 2.5 |
| 30 | mp-1215592 | $Zr_2CN$ | $R\bar{3}m$ | 4 | 2.0E-07 | -1.38 | 14.8 | 2.1 |
| 31 | mp-1217142 | $Ti_2CN$ | $R\bar{3}m$ | 4 | 0.0E+00 | -1.39 | 13.4 | 1.6 |
| 32 | mp-1219706 | $PrThN_2$ | P4/mmm | 4 | 1.0E-07 | -1.70 | 17.9 | 1.5 |
| 33 | mp-1215334 | $ZrAl_3N_4$ | Pmmm | 8 | 3.3E-04 | -1.36 | 10.1 | 1.4 |
| 34 | mp-1223850 | $Hf_2CN$ | $R\bar{3}m$ | 4 | 3.0E-07 | -1.50 | 11.8 | 1.2 |
| 35 | mp-1217025 | $TiAlN_2$ | $R\bar{3}m$ | 4 | 3.3E-03 | -1.60 | 13.9 | 1.2 |
| 36 | mp-1222707 | $LaThN_2$ | $R\bar{3}m$ | 4 | 2.0E-05 | -1.75 | 18.7 | 1.0 |
| 37 | mp-1226556 | $CeThN_2$ | $R\bar{3}m$ | 4 | 9.3E-05 | -1.76 | 17.9 | 0.9 |
| 38 | mp-1219257 | ScTaN | $P\bar{3}m1$ | 6 | 1.4E-04 | -1.14 | 12.0 | 0.7 |
| 39 | mp-9587 | TaMoN | P4/nmm | 6 | 6.2E-03 | -0.88 | 9.8 | 0.6 |
| 40 | mp-1215890 | $YThN_2$ | P4/mmm | 4 | 1.1E-06 | -1.86 | 17.0 | 0.5 |
| 41 | mp-1246326 | $Ti_2SeN_2$ | $P\bar{3}m1$ | 5 | 3.7E-04 | -1.51 | 12.4 | 0.4 |
| 42 | mp-1215389 | $Zr_4C_3N$ | $R\bar{3}m$ | 8 | 0.0E+00 | -1.10 | 12.3 | 0.4 |
| 43 | mp-1084767 | $Zr_2InN$ | $P6_3/mmc$ | 8 | 4.2E-06 | -1.24 | 9.6 | 0.4 |
| 44 | mp-1221519 | $Mo_2CN$ | Pmm2 | 4 | 4.0E-07 | -0.35 | 9.9 | 0.3 |
| 45 | mp-1018783 | LiBeB | $P2_1/m$ | 6 | 0.0E+00 | -0.12 | 17.5 | 0.3 |
| 46 | mp-1207086 | $MgAlB_4$ | P6/mmm | 6 | 2.0E-07 | -0.14 | 11.9 | 0.1 |
| 47 | mp-1245927 | $Ti_2SN_2$ | $P\bar{3}m1$ | 5 | 2.5E-04 | -1.58 | 12.7 | 0.0 |
| 48 | mp-1216653 | $UNbC_2$ | $R\bar{3}m$ | 4 | 1.5E-03 | -0.41 | 10.6 | 0.0 |
| 49 | mp-4978 | $Ti_2AlN$ | $P6_3/mmc$ | 8 | 0.0E+00 | -1.29 | 10.9 | 0.0 |
| 50 | mp-1215898 | $YScN_2$ | $R\bar{3}m$ | 4 | 5.0E-07 | -1.91 | 13.8 | 0.0 |
| 51 | mp-1215195 | $ZrUC_2$ | $R\bar{3}m$ | 4 | 2.6E-04 | -0.49 | 9.8 | 0.0 |
| 52 | mp-2701 | NbN | $P6_3/mmc$ | 4 | 0.0E+00 | -1.20 | 15.0 | 0.0 |
| 53 | mp-2305 | MoC | $P\bar{6}m2$ | 2 | 0.0E+00 | -0.08 | 9.8 | 0.0 |
| 54 | mp-570998 | $NaLi_2N$ | P6mmm | 4 | 4.0E-06 | -0.25 | 22.2 | 0.0 |
| 55 | mp-16726 | LiB | Pnma | 8 | 9.6E-06 | -0.15 | 20.7 | 0.0 |
| 56 | mp-2634 | NbN | $P\bar{6}m2$ | 2 | 0.0E+00 | -1.25 | 15.0 | 0.0 |
| 57 | mp-865 | $CaB_6$ | $Pm\bar{3}m$ | 7 | 0.0E+00 | -0.47 | 14.6 | 0.0 |
| 58 | mp-1208952 | $ScTaN_2$ | $P6_3/mmc$ | 8 | 0.0E+00 | -1.82 | 13.4 | 0.0 |
| 59 | mp-4262 | BeAlB | $F\bar{4}3m$ | 3 | 4.0E-07 | -0.05 | 11.8 | 0.0 |
| 60 | mp-1221393 | $MoWC_2$ | Pmm2 | 4 | 1.0E-07 | -0.10 | 11.2 | 0.0 |
| 61 | mp-954 | $BaB_6$ | $Pm\bar{3}m$ | 7 | 0.0E+00 | -0.41 | 10.1 | 0.0 |
| 62 | mp-242 | $SrB_6$ | $Pm\bar{3}m$ | 7 | 0.0E+00 | -0.52 | 10.1 | 0.0 |
| 63 | mp-31055 | $Sc_3InN$ | $Pm\bar{3}m$ | 5 | 4.9E-05 | -1.19 | 9.8 | 0.0 |
| 64 | mp-29928 | $Na_3Li_3N_2$ | Pm | 8 | 2.2E-05 | -0.08 | 23.0 | – |
| 65 | mp-1001835 | LiB | $P6_3/mmc$ | 4 | 2.0E-07 | -0.22 | 20.7 | – |
| 66 | mp-1216732 | $TiNbN_2$ | P4/mmm | 4 | 1.2E-04 | -1.48 | 15.7 | – |
| 67 | mp-1216799 | $TiNbN_2$ | $R\bar{3}m$ | 4 | 1.1E-04 | -1.51 | 15.7 | – |
| 68 | mp-15799 | NbN | $P6_3/mmc$ | 8 | 1.1E-03 | -1.04 | 15.0 | – |
| 69 | mp-1018142 | NbN | $F\bar{4}3m$ | 2 | 5.5E-03 | -0.83 | 15.0 | – |
| 70 | mp-1580 | NbN | $Fm\bar{3}m$ | 2 | 1.0E-04 | -1.04 | 15.0 | – |
| 71 | mp-1215256 | $ZrNbN_2$ | $R\bar{3}m$ | 4 | 4.4E-03 | -1.52 | 14.4 | – |
| 72 | mp-1217914 | $TaTiN_2$ | P4/mmm | 4 | 9.5E-04 | -1.53 | 14.0 | – |
| 73 | mp-1224296 | $HfTaN_2$ | $R\bar{3}m$ | 4 | 2.6E-03 | -1.53 | 13.7 | – |
| 74 | mp-1224419 | $HfNbN_2$ | $R\bar{3}m$ | 4 | 6.5E-05 | -1.56 | 13.3 | – |
| 75 | mp-1215288 | $ZrAlN_2$ | P4/mmm | 4 | 4.5E-03 | -1.35 | 13.1 | – |
| 76 | mp-1071766 | $ZrTa_2N_3$ | P6/mmm | 6 | 4.5E-03 | -1.23 | 13.0 | – |
| 77 | mp-1078349 | $La_2BiN$ | P4/nmm | 8 | 1.3E-05 | -1.36 | 12.9 | – |
| 78 | mp-1220406 | $NbMoC_2$ | $R\bar{3}m$ | 4 | 1.1E-03 | -0.12 | 12.3 | – |
| 79 | mp-1220438 | $NbMoN_2$ | $R\bar{3}m$ | 4 | 5.9E-03 | -0.59 | 12.2 | – |
| 80 | mp-1246102 | $ZrInN_2$ | P4/nmm | 8 | 6.1E-05 | -0.85 | 12.1 | – |
| 81 | mp-1018718 | $HfTa_2N_3$ | P6/mmm | 6 | 7.4E-04 | -1.16 | 11.7 | – |
| 82 | mp-7522 | $Th_2SbN_2$ | I4/mmm | 5 | 8.6E-06 | -1.77 | 10.9 | – |
| 83 | mp-972309 | $Ta_2BN_3$ | P6/mmm | 6 | 6.2E-06 | -0.25 | 10.9 | – |
| 84 | mp-1226227 | $CrB_4Mo$ | Immm | 6 | 3.3E-03 | -0.25 | 10.8 | – |
| 85 | mp-1077140 | $Zr_2TaN_3$ | P6/mmm | 6 | 9.1E-03 | -1.14 | 10.7 | – |
| 86 | mp-27469 | $Th_2BiN_2$ | I4/mmm | 5 | 4.8E-05 | -1.67 | 10.7 | – |
| 87 | mp-10675 | $Ti_3AlN$ | $Pm\bar{3}m$ | 5 | 8.0E-05 | -1.03 | 10.7 | – |
| 88 | mp-1217826 | $TaWC_2$ | $R\bar{3}m$ | 4 | 2.7E-04 | -0.14 | 10.6 | – |
| 89 | mp-1029446 | $CaGeN_2$ | P4/nmm | 8 | 0.0E+00 | -0.60 | 10.4 | – |
| 90 | mp-1078784 | $Sc_3RhC_4$ | Immm | 8 | 1.0E-07 | -0.49 | 10.3 | – |
| 91 | mp-1215903 | $YBN_2$ | $P\bar{4}m2$ | 4 | 1.2E-04 | -0.50 | 10.2 | – |
| 92 | mp-1018710 | $Hf_2TaN_3$ | P6/mmm | 6 | 1.9E-04 | -1.27 | 10.2 | – |
| 93 | mp-8996 | $Sc_3RuC_4$ | Immm | 8 | 0.0E+00 | -0.47 | 10.0 | – |
| 94 | mp-1226555 | $CeThC_2$ | $R\bar{3}m$ | 4 | 3.7E-04 | -0.16 | 9.9 | – |
| 95 | mp-1019273 | $TaTi_2N_3$ | P6/mmm | 6 | 4.1E-04 | -1.22 | 9.9 | – |
| 96 | mp-1217919 | $TaMoC_2$ | $R\bar{3}m$ | 4 | 4.9E-04 | -0.18 | 9.9 | – |
| 97 | mp-1217822 | $NbVC_2$ | $R\bar{3}m$ | 4 | 5.0E-05 | -0.46 | 9.9 | – |
| 98 | mp-1217903 | $TaMo_2C_3$ | $P\bar{3}m1$ | 6 | 1.9E-03 | -0.05 | 9.7 | – |
| 99 | mp-1217764 | $Ta_2CN$ | P4/mmm | 4 | 7.4E-03 | -0.81 | 9.6 | − |
| 100 | mp-1216711 | $TiMoC_2$ | $R\bar{3}m$ | 4 | 2.9E-04 | -0.40 | 9.5 | − |

## 3. Phonon spectra of compounds listed in Table III of the main text

We calculate the phonon spectra of all the compounds listed in Table. III of the main text. Each figure contains the Materials Project ID number (mp-id), electron-phonon coupling strength ($\lambda_{qv}$), and the Eliashberg spectral function $\alpha^2F(\omega)$ and $\lambda(\omega)$. These figures are presented in Figs. S5 (a–h), Figs. S6 (i–p), and Figs. S7 (q–r).

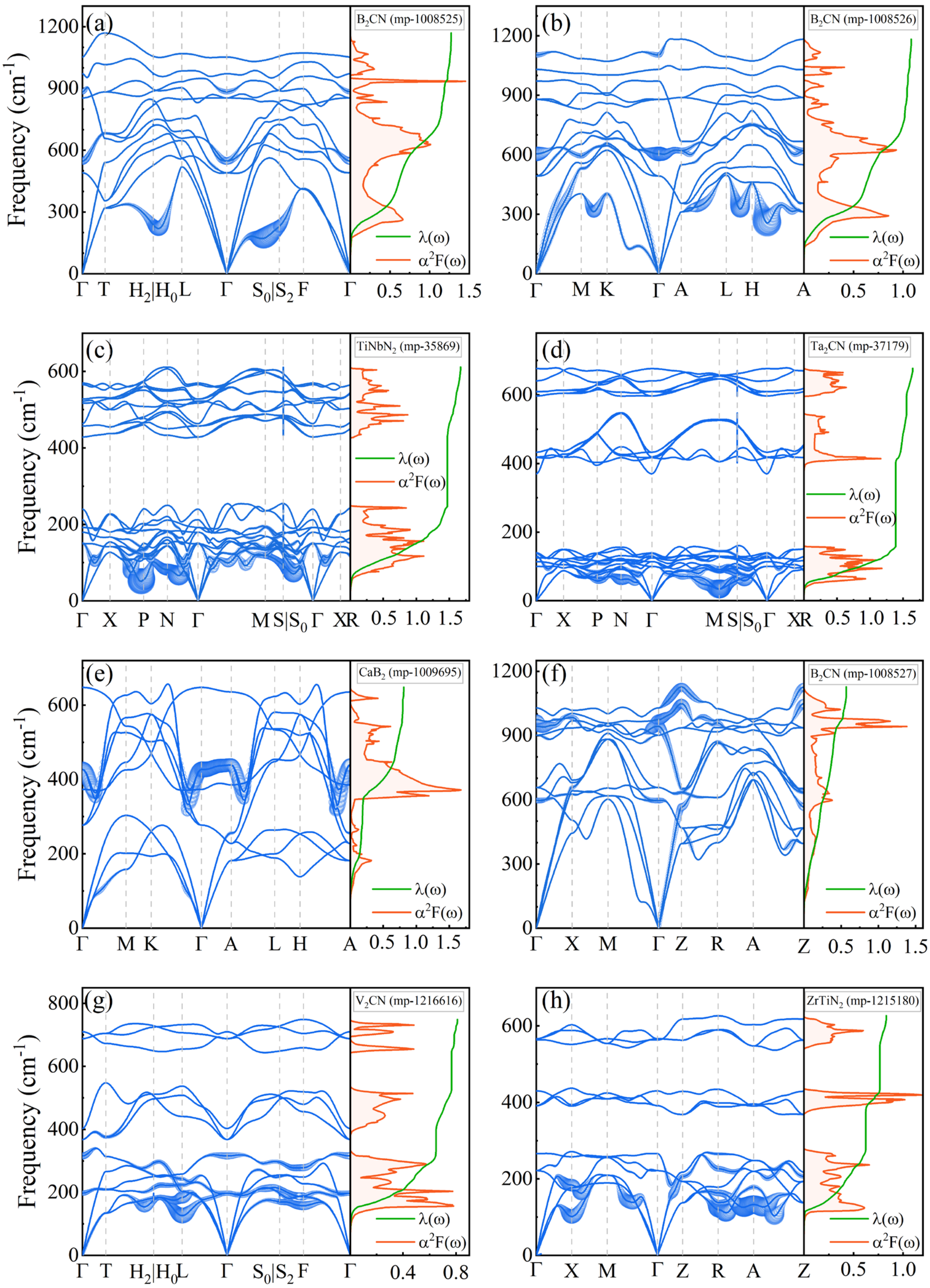

FIG. S5. Phonon spectra (left) and Eliashberg spectral function $\alpha^2F(\omega)$ and $\lambda(\omega)$ (right).

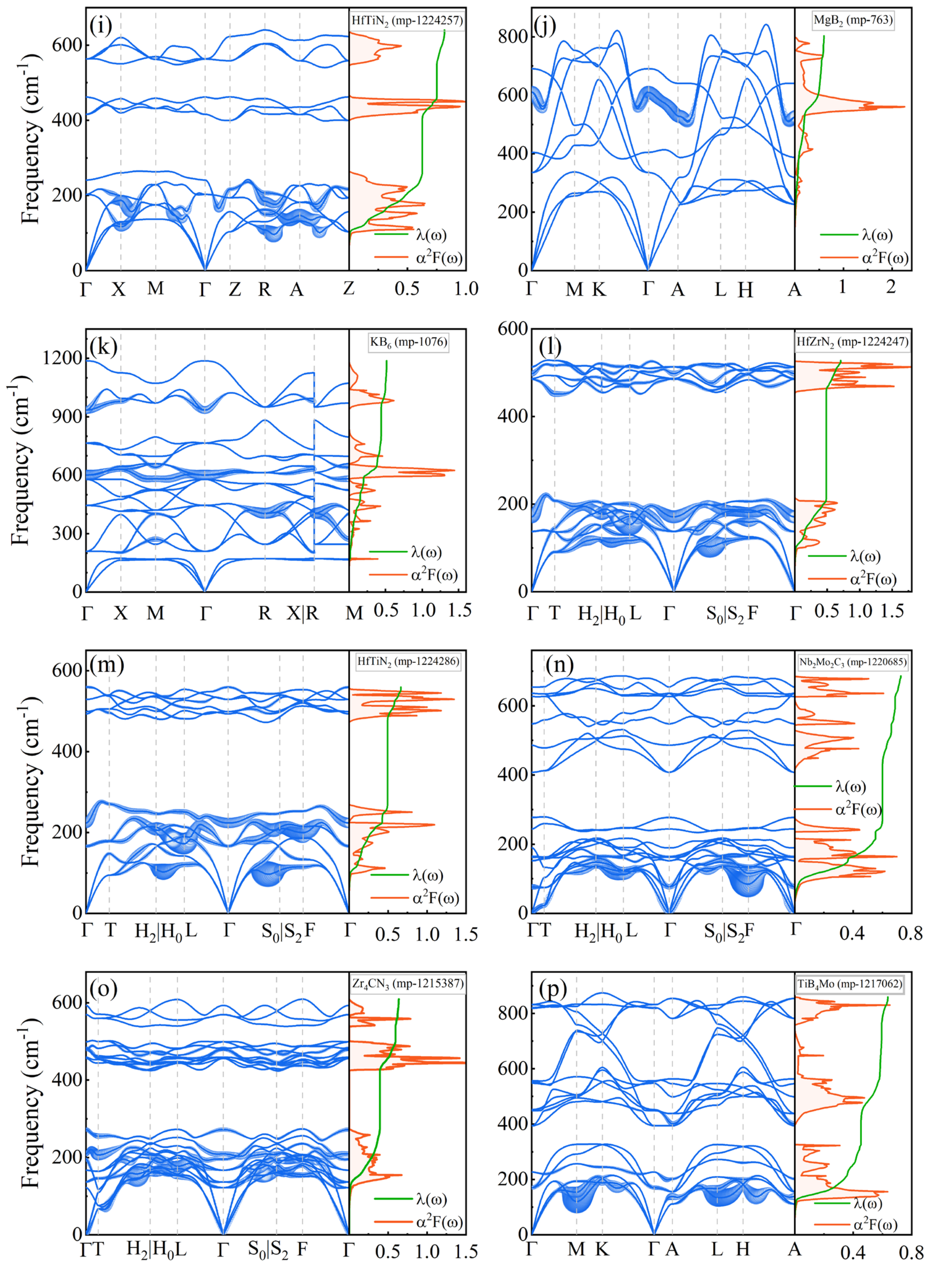

FIG. S6. Phonon spectra (left) and Eliashberg spectral function $\alpha^2F(\omega)$ and $\lambda(\omega)$ (right).

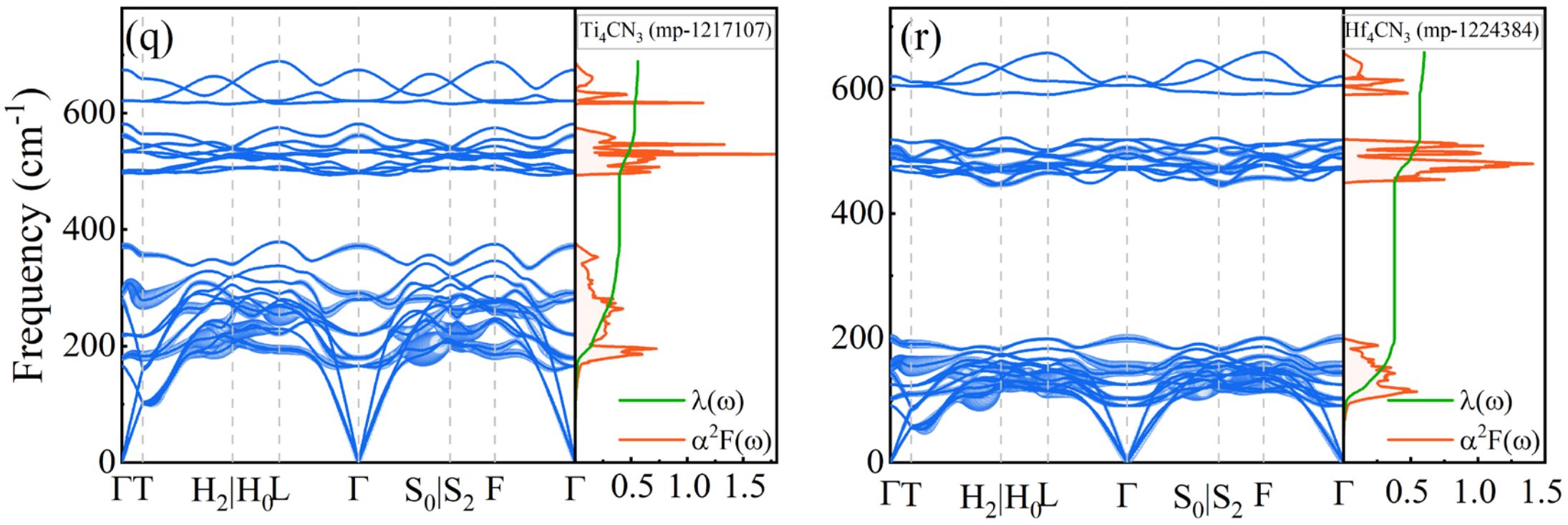

FIG. S7. Phonon spectra (left) and Eliashberg spectral function $\alpha^2F(\omega)$ and $\lambda(\omega)$ (right).

## 4. Electronic structures of $R_3Ni_2O_7$

We calculated the electronic structures of $R_3Ni_2O_7$, including orbital-projected band structures and density of states under 0 GPa (Figs. S8 and Figs. S9) and 30 GPa (Figs. S10 and Figs. S11), and Fermi surfaces under 0 and 30 GPa (Figs. S12 and Figs. S13).

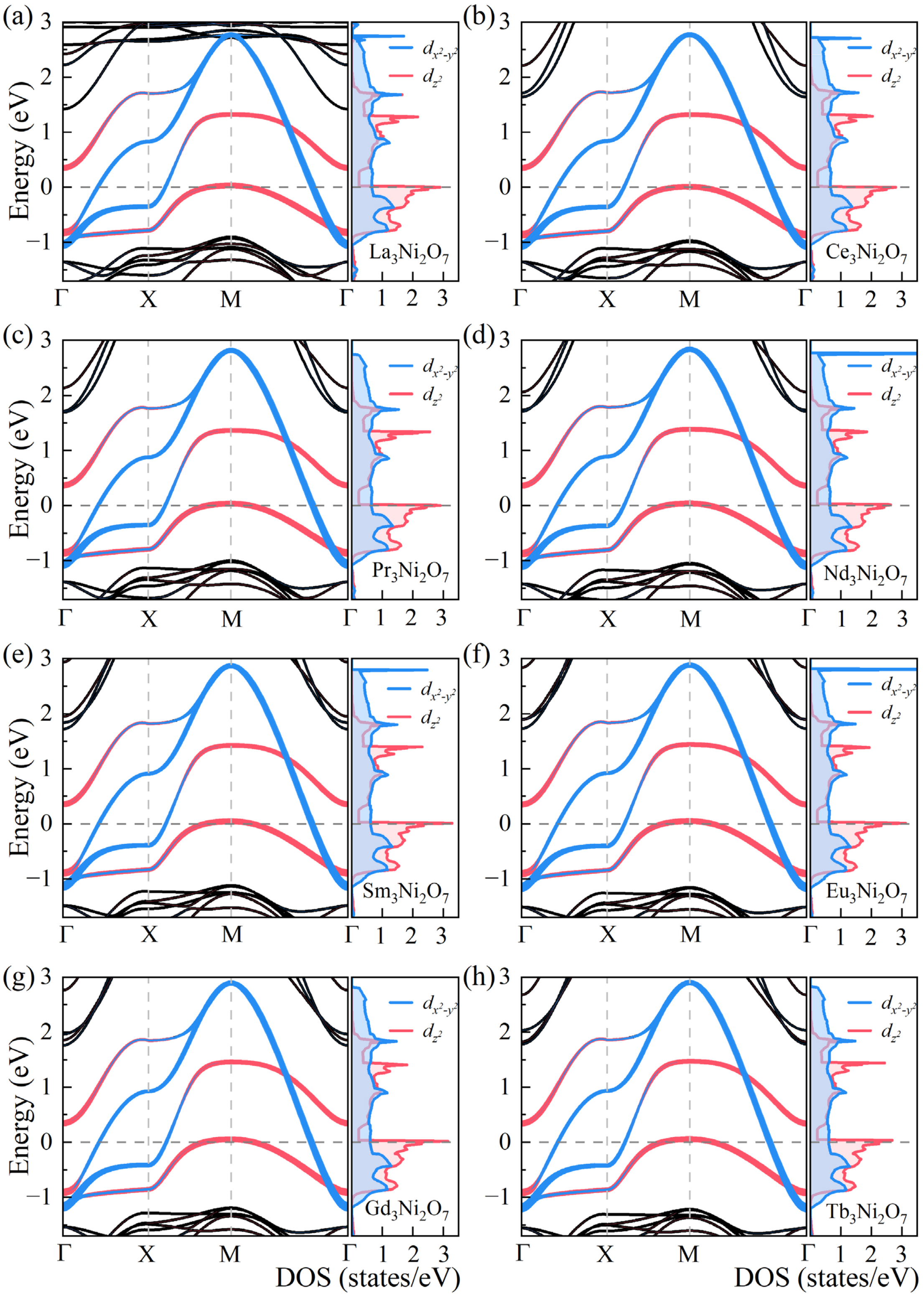

FIG. S8. Orbital-projected band structure and density of states of $R_3Ni_2O_7$ (R = La ~ Tb) under 0 GPa.

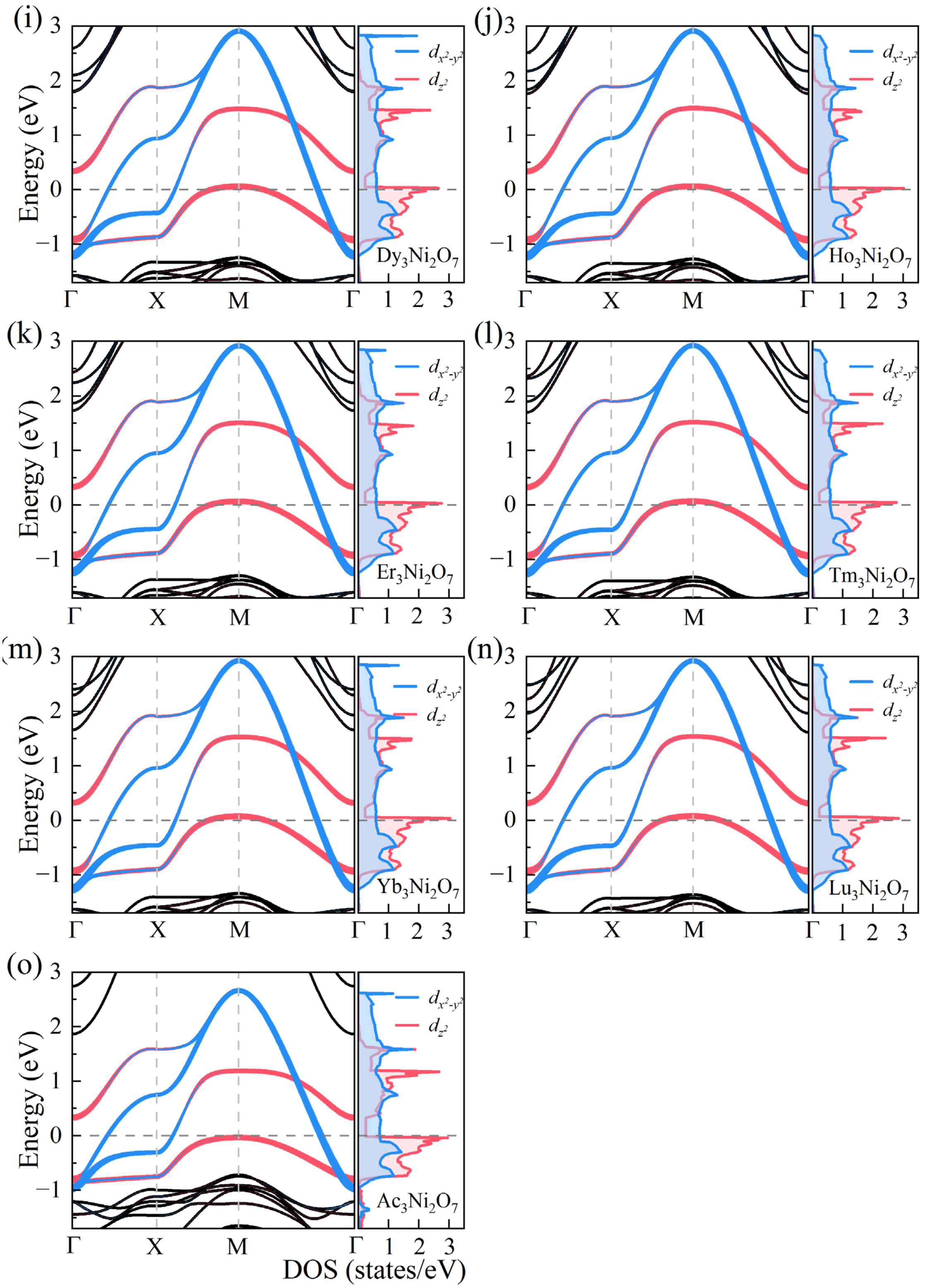

FIG. S9. Orbital-projected band structure and density of states of $R_3Ni_2O_7$ (R = Dy ~ Ac) under 0 GPa.

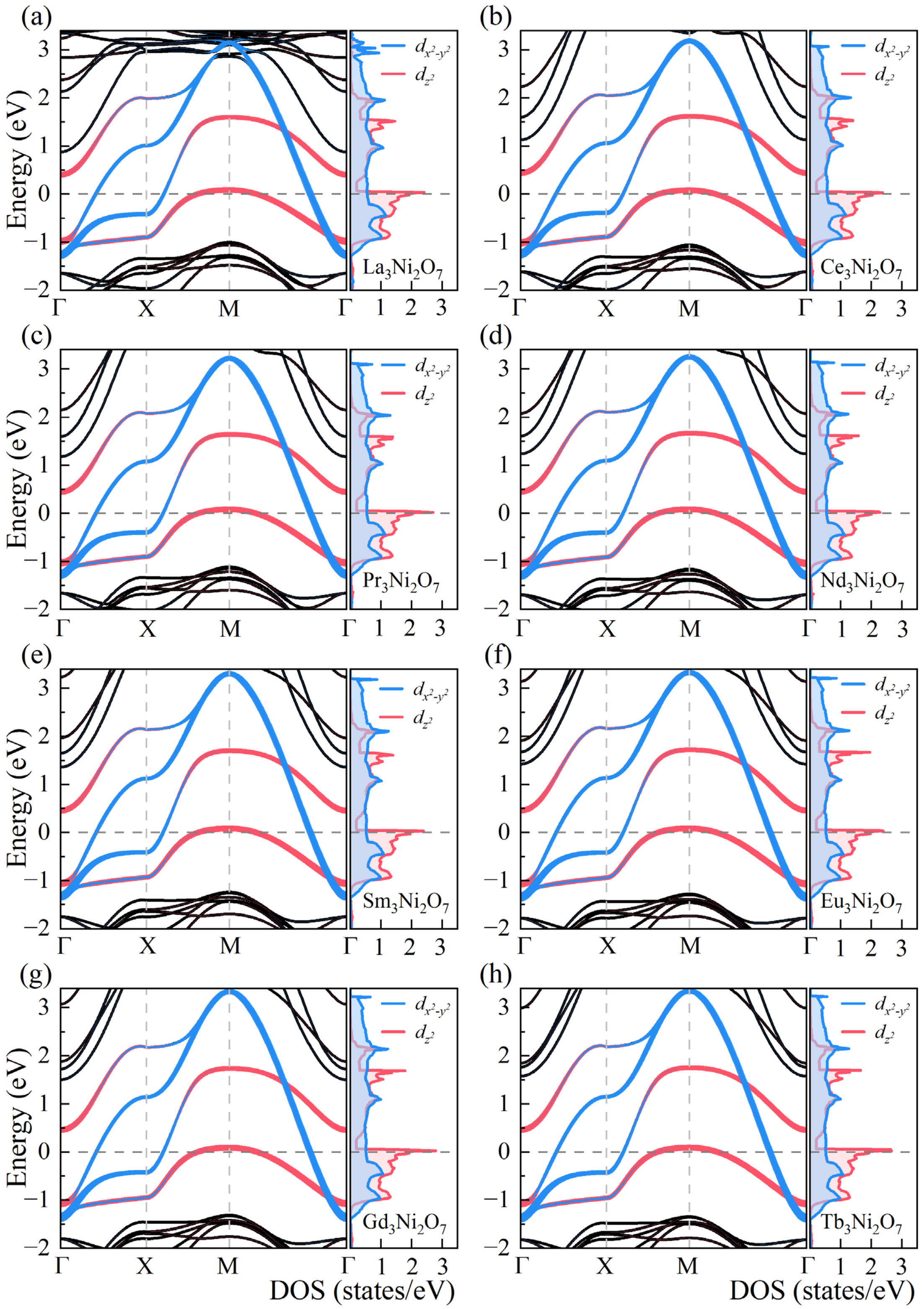

FIG. S10. Orbital-projected band structure and density of states of $R_3Ni_2O_7$ (R = La ~ Tb) under 30 GPa.

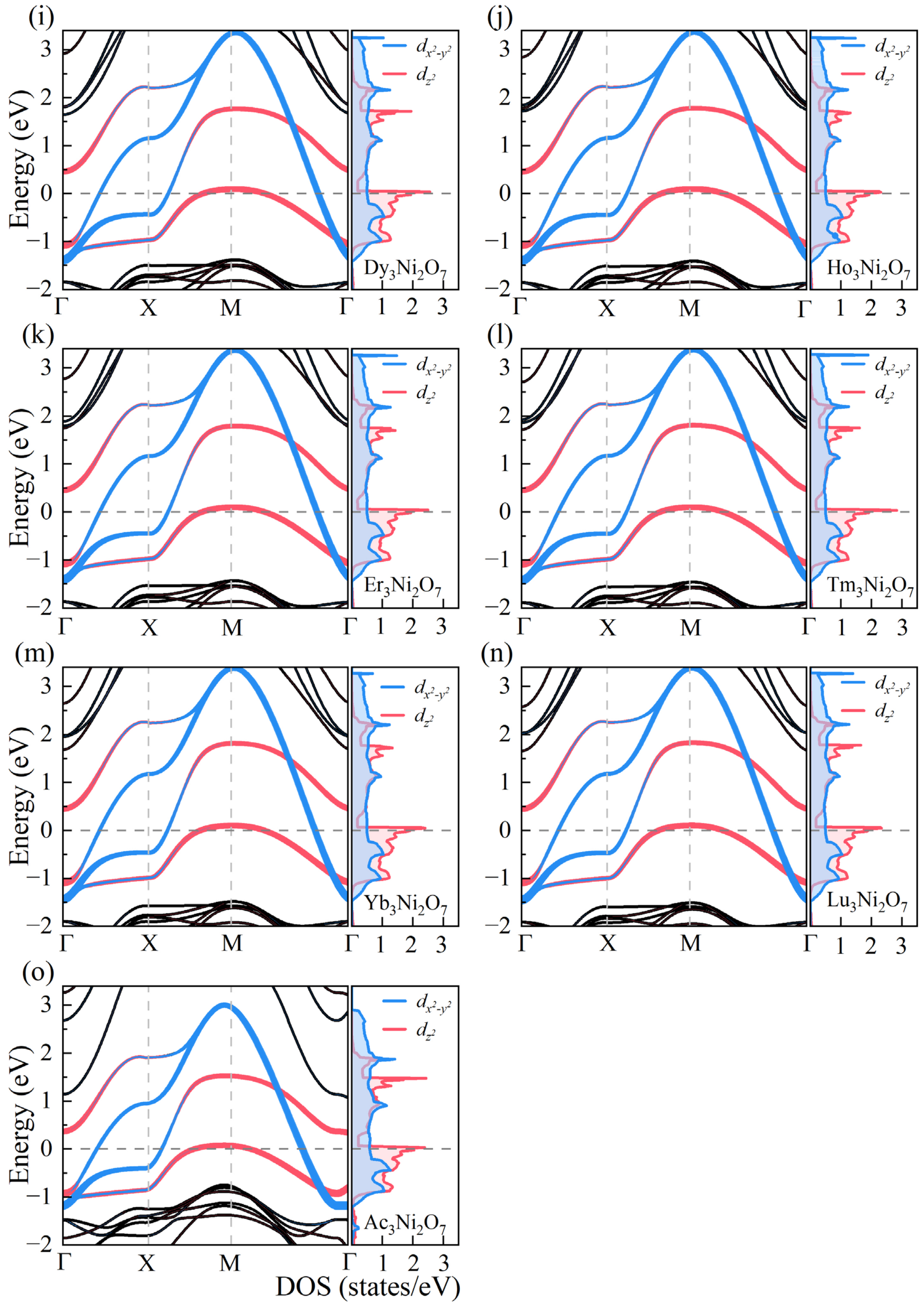

FIG. S11. Orbital-projected band structure and density of states of $R_3Ni_2O_7$ (R = Dy ~ Ac) under 30 GPa.

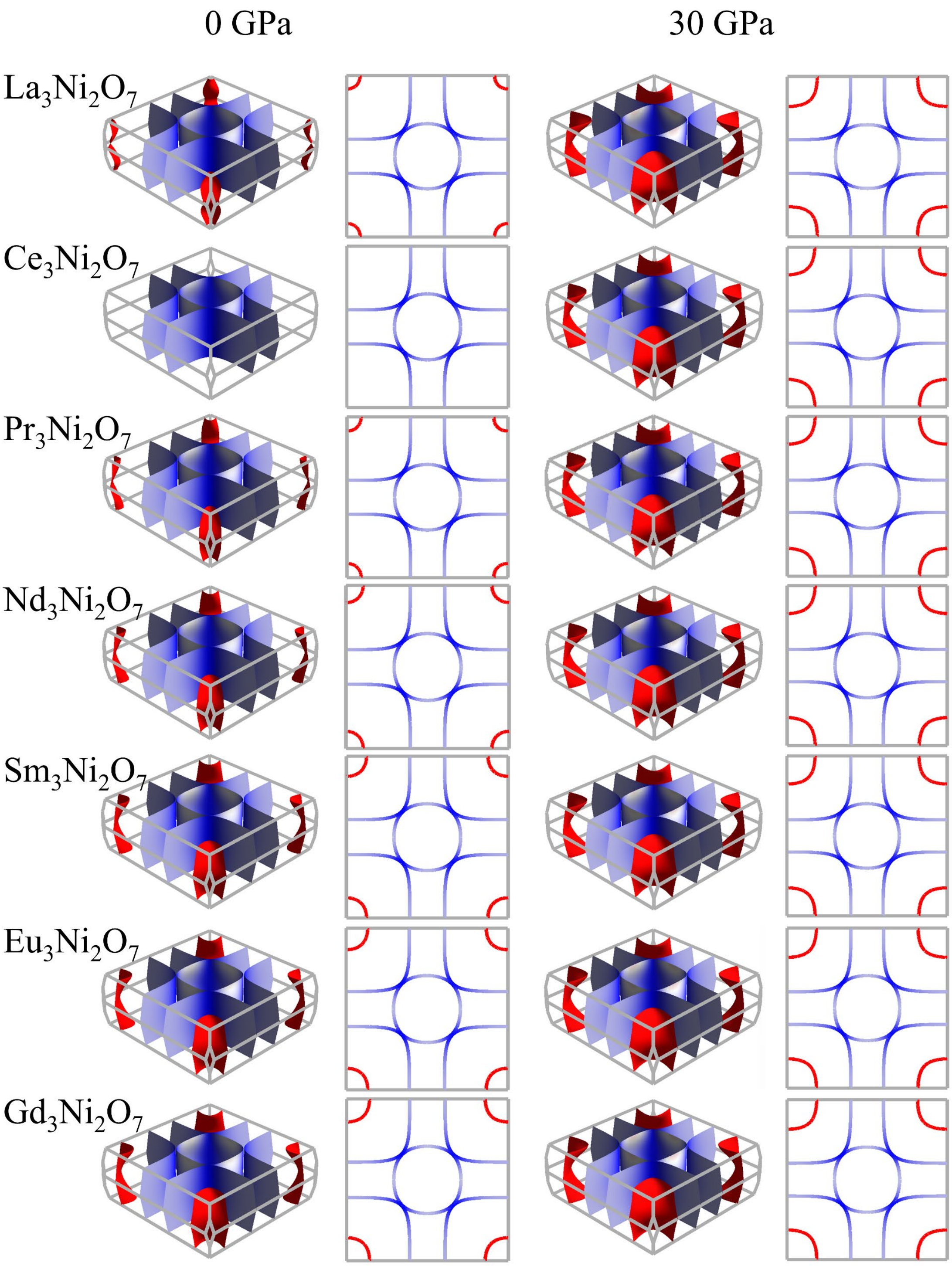

FIG. S12. Fermi surfaces in the three-dimensional Brillouin zone and projection on two dimensions with orbital weights of $R_3Ni_2O_7$ (R = La ~ Gd) under 0 and 30 GPa.

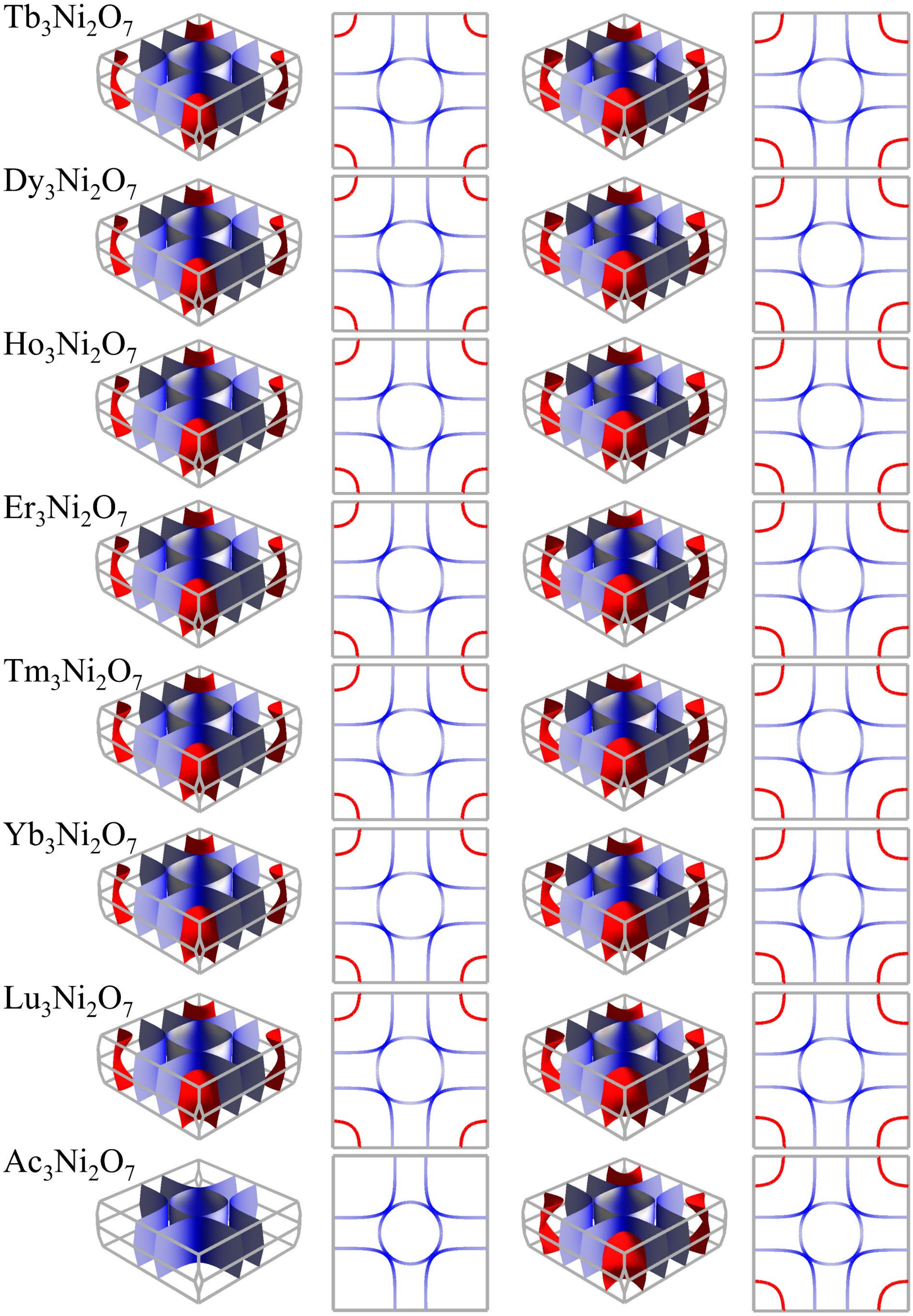

FIG. S13. Fermi surfaces in the three-dimensional Brillouin zone and projection on two dimensions with orbital weights of $R_3Ni_2O_7$ (R = Tb ~ Ac) under 0 and 30 GPa.